\documentclass[twocolumn,aps,groupedaddress,nopacs,footinbib,preprintnumbers,floats,floatfix]{revtex4}
\usepackage{graphicx}
\usepackage{verbatim}
\usepackage{ulem}
\usepackage{color}
\usepackage{float}
\usepackage{mathrsfs}
\usepackage{amssymb}
\usepackage{amsmath}
\usepackage{psfrag}
\usepackage{bm}
\usepackage{subfigure}%
\newcommand{\bra}[1]{\left(#1\right)}
\newcommand{\bras}[1]{\left[#1\right]}
\newcommand{\brac}[1]{\left\{#1\right\}}
\newcommand{\ber}{\begin{eqnarray}}
\newcommand{\eer}{\end{eqnarray}}

\newcommand{\be}{\begin{equation}}
\newcommand{\ee}{\end{equation}}
\newcommand{\ba}{\begin{eqnarray}}
\newcommand{\ea}{\end{eqnarray}}

\newcommand{\planck}{{\sc Planck}} 
\newcommand{\ADEPT}{{\sc Adept}}
\newcommand{\BOSS}{{\sc Boss}}

\newcommand{\adniv}{$\langle $\sc ad,niv$\rangle$}
\newcommand{\adnid}{$\langle $\sc ad,nid$\rangle$}
\newcommand{\adnivE}{\langle \mbox{\sc ad,niv}\rangle}
\newcommand{\adnidE}{\langle \mbox{\sc ad,nid}\rangle}

\newcommand{\adE}{\langle \mbox{\sc ad,ad}\rangle}

\newcommand{\nidE}{\langle \mbox{\sc nid,nid}\rangle}

\newcommand{\ad}{$\langle $\sc ad,ad$\rangle$}
\newcommand{\ci}{$\langle $\sc ci,ci$\rangle$}
\newcommand{\nid}{$\langle $\sc nid,nid$\rangle$}
\newcommand{\niv}{$\langle $\sc niv,niv$\rangle$}

\newcommand{\adm}{\sc ad}
\newcommand{\cim}{\sc ci}
\newcommand{\nidm}{\sc nid}
\newcommand{\nivm}{\sc niv}

\newcommand{\adci}{$\langle $\sc ad,ci$\rangle$}
\newcommand{\cinid}{$\langle $\sc ci,nid$\rangle$}
\newcommand{\ciniv}{$\langle $\sc ci,niv$\rangle$}
\newcommand{\nidniv}{$\langle $\sc nid,niv$\rangle$}

\newcommand{\adciE}{\langle \mbox{\sc ad,ci}\rangle}

\newcommand{\cinivE}{\langle \mbox{\sc ci,niv}\rangle}
\newcommand{\ISO}{\mbox{\sc iso}}

\newcommand{\Heav}{\text{H}}
\newcommand{\bftheta}{\mbox{\boldmath$\theta$}}
\def\gtorder{\mathrel{\raise.3ex\hbox{$>$}\mkern-14mu
             \lower0.6ex\hbox{$\sim$}}}

\def\l{{\cal L}}

\begin{document}

\title{The Sensitivity of BAO Dark Energy Constraints to General Isocurvature Perturbations}
\author{S. Muya Kasanda$^{1,4}$, C. Zunckel$^{1,2}$, K. Moodley$^{1,4}$ $\&$ B. A. Bassett$^{3,4,5,6}$,  P. Okouma$^{3,4,5}$\\
\it $^1$ Astrophysics and Cosmology Research Unit, University of KwaZulu-Natal, Durban, 4041, SA\\
\it $^2$ Astrophysics Department, Princeton University, Peyton Hall, 4 Ivy Lane, NJ, 08544, USA\\
\it $^3$ Dept. of Mathematics and Applied Mathematics, University of Cape Town, Rondebosch 7701, Cape Town, SA\\\it $^4$ Centre for High Performance Computing, CSIR Campus, 15 Lower Hope St., Rosebank, Cape Town, SA\\\it $^5$ African Institute for Mathematical Sciences, Muizenberg, Cape Town, SA\\\it $^6$ South African Astronomical Observatory, P.O. Box 9, Observatory 7935, SA}

\begin{abstract}
Baryon Acoustic Oscillation (BAO) surveys will be a leading method for addressing the dark energy challenge in the next decade. We explore in detail the effect of allowing for small amplitude admixtures of general isocurvature perturbations in addition to the dominant adiabatic mode. We find that non-adiabatic initial conditions leave the sound speed unchanged but instead excite different harmonics. These harmonics couple differently to Silk damping, altering the form and evolution of acoustic waves in the baryon-photon fluid prior to decoupling. This modifies not only the scale on which the sound waves imprint onto the baryon distribution, which is used as the standard ruler in BAO surveys, but also the shape, width and height of the BAO peak. We discuss these effects in detail and show how more general initial conditions impact our interpretation of cosmological data in dark energy studies. We find that the inclusion of these additional isocurvature modes leads to a decrease in the Dark Energy Task Force figure of merit (FoM) by $46\%$ i.e., $\mbox{FoM}_{ISO}=0.54\times\mbox{FoM}_{AD}$ and $53\%$ for the {\BOSS} and {\ADEPT} experiments respectively when considered in conjunction with {\planck} data. We also show that the incorrect assumption of adiabaticity has the potential to bias our estimates of the dark energy parameters by $2.7\sigma$ ($2.2\sigma$) for a single correlated isocurvature mode (CDM isocurvature), and up to $4.9\sigma$ ($5.7\sigma$) for three correlated isocurvature modes in the case of the {\BOSS} ({\ADEPT}) experiment. We find that the use of the large scale structure data in conjunction with CMB data improves our ability to measure the contributions of different modes to the initial conditions by as much as $95\%$ for certain modes in the fully correlated case. 

\end{abstract}

\maketitle

\section{Introduction}

Although the standard model of cosmology based on $\Lambda$CDM has not changed fundamentally in the last decade, there has been a remarkable refinement in our knowledge of the parameters describing the model. For example, the original supernova results gave only limits of $\Omega_m < 1.5$ at $2\sigma$ assuming a general $\Lambda$CDM model \cite{earlysn} while the latest results from the WiggleZ Baryon Acoustic Oscillation (BAO) survey, together with WMAP and Union2 supernova data now give $\Omega_m = 0.29 \pm 0.04$ at $2\sigma$ \cite{blake11}.

As a result of this progress it has become obvious that systematic errors are a key issue in pushing the frontier further. For example, in the case of supernovae there are important systematic errors related to the lightcurve fitter used which currently leads to shifts in the dark energy equation of state of about $\Delta w \sim 0.1$ \cite{kessler}. There may be additional supernova systematics such as the existence of Type Ia subpopulations and correlations between absolute magnitude and host galaxy type (see e.g. \cite{subpops}).

BAO have their own associated systematic errors, such as nonlinear effects which potentially bias or shift the BAO peak, although these are believed to be fairly small and possible to calibrate through theoretical modeling and N-body simulations \cite{baonl}. However there is another theoretical systematic due to isocurvature perturbations that has recently received attention \cite{ibao1,ibao2}. Depending on how general one allows the primordial isocurvature admixture to be, there can be a significant impact on the ability of future BAO surveys to constrain dark energy even if one imposes the constraint that the isocurvature modes be undetectable by {\planck} alone \cite{ibao1}. This shows that at least in the next generation of surveys one will not be able to decouple the search for dark energy with BAO from an understanding of the early universe, a subtlety that does not affect supernovae surveys.

The key reason that even small correlated isocurvature modes cause a problem for BAO surveys is that they alter the way in which the BAO peak appears in the two-point correlation function of baryons, and hence, of galaxies. In the simple adiabatic model the BAO peak is controlled by the sound horizon, the distance that sound waves can propagate in the early universe from the time of inflation to decoupling.  This characteristic scale depends only on the sound speed in the standard adiabatic picture:
\be
r_{s} = \int^{t_{cmb}}_0 c_s(1+z) dt = \int^{\infty}_{z_{cmb}} \frac{c_s(z')}{H(z')} dz'
\label{soundhorizon}
\ee where $c_s(z)=1/\sqrt{3\bra{1+R_b/(1+z)}}$ and $R_b=31500\mbox{ }\omega_b\bra{T_{\mbox{cmb}}/2.7\mbox{ K}}^{-4}$ which can be measured accurately with the Cosmic Microwave Background (CMB).  By comparing the size of this scale at the time of decoupling and its angular size at late times we can learn about the expansion history of the universe and measure cosmic distances, and hence constrain models of dark energy \cite{BAOReviews}.  

In \cite{ibao1}, we found a clear degeneracy between the impact of dark energy models and non-adiabatic initial conditions on the galaxy correlation function.   In this paper we explain in depth why small amplitude but general admixtures of correlated isocurvature modes can have such a strong impact on the cosmological constraints based on BAO surveys. We show that relaxing the assumption of adiabaticity and allowing fractions of isocurvature modes affects the development of the acoustic waves in the baryon-photon fluid. The isocurvature modes excite different harmonics which in turn, couple differently to Silk damping,  and in so doing, modifies both the scale on which the sound waves imprint on the baryon distribution and the shape of the BAO peak.

This paper is arranged as follows; in section \ref{sec:BAP}, we study the evolution of the baryon density contrast under different initial conditions and how the structure of the BAO peak is altered. In section \ref{sec:FM}, a Fisher matrix formalism is implemented in order to quantify the impact of these changes on the forecasted errors on the dark energy parameters from two BAO experiments, namely {\BOSS} \cite{Boss} and {\ADEPT} \cite{ADEPT}.  As a prior, we include the information from the high-resolution CMB temperature anisotropy and polarization spectra from the {\planck} Surveyor \cite{Planck}, which should provide stringent constraints on the amount of isocurvature in the initial conditions. We also conduct a study of the potential bias in our estimates of the dark energy parameters that can result from an incorrect assumption of pure adiabatic initial conditions. Lastly, we show that constraints on the isocurvature parameters can be derived from BAO surveys. We discuss our conclusions in section \ref{sec:conclusion}.

\section{The BAO peak with adiabatic and isocurvature initial conditions}
\label{sec:BAP}
\noindent
 
The BAO peak is sensitive not only to the matter content of the universe, but also to the character of the 
primordial perturbations. The features of the BAO peak such as the location, width and amplitude are mainly dictated by the time evolution of the baryon density contrast $\delta_b$ from the post-inflation period to photon-baryon decoupling. In turn, the time evolution of the baryon density contrast during the pre-decoupling period depends on the initial configuration of the primordial perturbations in the different species at the end of inflation.\\

In the simplest scenario the perturbation affects all the cosmological species such that the relative ratios in the number densities remain unperturbed, exciting the adiabatic mode (AD). Although adiabatic initial conditions are a natural feature of single-field inflationary models \cite{single_field}, it has been shown \cite{Bucher00} that four regular isocurvature (ISO) modes are allowed in addition to the adiabatic (AD) mode. These isocurvature modes are characterized by variations in the particle number ratios but with vanishing curvature perturbation, with different isocurvature modes excited depending on the species that are initially perturbed. These are namely the cold dark matter isocurvature (CI) mode, the baryon isocurvature (BI) mode, the neutrino isocurvature density (NID) and the neutrino isocurvature velocity (NIV) mode. While isocurvature modes are more difficult to physically motivate, the possibility of correlated isocurvature fluctuations is allowed given current cosmological data \cite{Bean06,Bucher01} and  interesting to consider.

We will show that different modes of the primordial perturbations excite different harmonics and these harmonics couple differently to the Silk damping, thereby altering the characteristic BAO scale at photon-baryon decoupling. After decoupling, baryon fluctuations on scales larger than the Jeans length $\lambda_J$ slow down in the rest frame of the cold dark matter (CDM), falling into the CDM potential wells, and eventually tracing the CDM, while on scales below $\lambda_J$ the fluctuations still oscillate, independently of the initial conditions.  In order to study the features of the BAO peak for different modes, we consider the time evolution of the photon-baryon fluid in the tight-coupling regime.\\

\noindent

In this regime, photons and baryons are treated as perfect fluids. The subscripts $b$, $c$, $\gamma$ and $\nu$ respectively denote the baryons, CDM, photons and neutrinos. The conservation of energy-momentum leads to the following set of time evolution equations for the photon and the baryon density contrasts $\delta$ and velocity divergences $\theta$ in the synchronous gauge \cite{Ma_Bertschinger_1995}:
\begin{align}
\label{delta_g}
\dot{\delta}_{\gamma}&=-\frac{4}{3}{\theta}_{\gamma}-\frac{2}{3}\dot{h},\\
\label{delta_b}
\dot{\delta}_{b}&=-\theta_{b}-\frac{1}{2}\dot{h},
\end{align}
for the density contrasts, and
\begin{align}
\label{theta_g}
\dot{\theta}_{\gamma}&=k^2\left(\frac{1}{4}\delta{\gamma}-\sigma_{\gamma}\right)+a n_e \sigma_T(\theta_b-\theta_{\gamma}),\\
\label{theta_b}
\dot{\theta}_b&=-\frac{\dot{a}}{a}\theta_b+c_s^2 k^2\delta_b+\frac{4\bar{\rho}_{\gamma}}{3\bar{\rho}_{b}}a n_e\sigma_T(\theta_{\gamma}-\theta_b),
\end{align}
for the velocity divergences. Here and throughout the paper, $\sigma_T$ is the Thomson scattering cross section, $n_e$ is the electron number density, $a$ is the scale factor, $\bar{\rho}$ is the background  density, $c_s$ is the sound speed given by $c_s=1/\sqrt{3(1+R)}$, $R=\frac{3\bar{\rho}_b}{4\bar{\rho}_{\gamma}}$ is the baryon-to-photon density ratio, $\sigma_{\gamma}$ is the photon shear, and the dot refers to the derivative with respect to the conformal time $\tau$. The variable $h$ is the metric field in synchronous gauge, which evolves according to \cite{Ma_Bertschinger_1995}
\begin{equation}
\label{metric_field}
\ddot{h}+\frac{\dot{a}}{a}h=-3\left(\frac{\dot{a}}{a}\right)^2\bar{\rho}_{cr}\sum_j\Omega_j\delta_j(1+3{{c_s}_j}^2),
\end{equation}
where $j\in \{\nu,\gamma,b,c\}$ labels the different species, $\bar{\rho}_{cr}$ is the critical density of the universe and $\Omega_j\equiv\bar{\rho}_j/\bar{\rho}_{cr}$ is the ratio of the density of the $j^{th}$ species to the critical density.

The tight-coupling approximation allows us to set $\theta_{\gamma}=\theta_b=\theta_{\gamma b}$, with the photon-baryon velocity evolving as
\begin{equation}
\label{eqn03_prior}
(1+R)\dot{\theta}_{\gamma b}=-\dot{R}\theta_{\gamma b}+k^2(\frac{1}{4}\delta_{\gamma}-\sigma_{\gamma})+c_s^2 k^2 R \delta_b,
\end{equation}
and the photon density contrast evolving as \cite{Hu_Sugiyama_1995}
\begin{equation}
\label{diff1}
\ddot{\delta}_{\gamma} +\frac{\dot{R}}{1+R}\dot{\delta}_{\gamma}+k^2c_s^2\delta_{\gamma}=-\frac{2}{3}\left[\frac{\dot{
R}}{1+R}\dot{h}+\ddot{h}\right].
\end{equation}
Here, we have neglected the photon shear (tight-coupling regime) and the pressure term in $\delta_b$ as it remains  smaller than the term in $\delta_{\gamma}$ prior to decoupling. Equation (\ref{diff1}) represents a driven harmonic oscillator with the competition between gravitational infall and photon pressure giving rise to acoustic waves propagating in the photon-baryon fluid at the speed of sound.

For the associated homogeneous equation, we look for solutions of the form $\delta_{\gamma}\propto\exp{i\int_0^{\tau}\omega d\tau'}$ where $\omega(\tau)$ is some phase function. The  two solutions to the homogeneous equation are simply $\sin{kr_s}$ and $\cos{kr_s}$, where $r_s(\tau)=\int_0^{\tau}c_s d\tau'$, the phase function is $\omega=kc_s$, and we have made use of the WKB approximation. On large scales, the WKB approximation breaks down, but these modes are irrelevant for the BAO treatment as they only enter the horizon well after decoupling. The particular solution is constructed by integrating the driving term weighted by the Green's function of the two homogeneous solutions \cite{Hu_Sugiyama_1995}. Thus, the time evolution of the acoustic waves in the photon component for all initial conditions prior to decoupling is given by
\begin{align}
\nonumber
&(1+R)^{1/2}\delta_{\gamma}(k,\tau)= A_S \sin{k r_s(\tau)} + A_C \cos{k r_s(\tau)}\\
\label{deltag}
&+\frac{1}{kc_s}\int_0^{\tau} (1+R(\tau'))^{1/2}\sin{[kr_s(\tau)-kr_s(\tau')]}F(\tau')d\tau',
\end{align}
where $A_S$ and $A_C$ are determined by the initial conditions as described in \cite{Bucher00}, and
\begin{equation}
\label{forcing}
F(\tau)=-\frac{2}{3}\left(\frac{\dot{R}}{1+R}\dot{h}+\ddot{h}\right),
\end{equation}
is the gravitational driving term which evolves differently for different initial conditions. Equation (\ref{deltag}) gives the time evolution of the photon density contrast in the tight-coupling regime. In this regime, the baryon density contrast is related to its photon counterpart by $\dot{\delta}_{b}=\frac{3}{4}\dot{\delta}_{\gamma}$. On small scales, a correction to the tight-coupling approximation must be applied when the Silk damping becomes important, as photons leak out of overdense regions, dragging baryons with them. This is done by multiplying the solution above by $e^{-k^2/k_D^2}$, where the photon diffusion scale $k_D^{-1}$ is given by $${k_D}^{-2}=\frac{1}{6}\int \frac{1}{\dot{\tau}_e}\frac{R^2+4(1+R)/5}{(1+R)^2},$$ where $\dot{\tau}_e=a n_e\sigma_T$ is the  differential optical depth. The Silk damping turns out to significantly affect both the shape and peak location of the BAO as we shall discuss later. After decoupling, the photons free stream, while baryons fall into the CDM potential wells under gravitational instability. Here, we only consider baryon fluctuations with wavelength larger than the Jeans scale.\\

The above description of the density contrast evolution in $k$-space can be intuitively and simply understood by looking at the evolution of the mass profile in the configuration space \cite{Eisenstein07,Bashinsky01}. The radial mass profile $M_j$ of a species $j$, given by
\begin{align}
M_{j}(r,z)&=\int_0^{\infty} T_{j}(k,z) \frac{\sin{kr}}{kr} k^2 r^2 dk,\\
\label{eq1}
&=r\int_0^{\infty} \delta_j(k,z)  \frac{\sin{kr}}{k} dk,
\end{align}
where $T_{j}(k,z)=\delta_j(k,z)/k^2$ is the transfer function of the $j^{th}$ species, describes the redshift evolution of a point-like overdensity initially located at the origin. The location of the mass profile peak gives the physical radius of the spherical shell of the  overdensity for a given species. For numerical computations, a Gaussian overdensity of width $\sigma^{-1}$ is used instead of a point-like overdensity. This is done by multiplying the integrand of equation (\ref{eq1}) by $e^{-k^2\sigma^2/2}$.\\

\noindent
Hereafter, we study the time evolution of the baryon mass profile for each mode in turn. We start from the well studied adiabatic case then move onto the isocurvature modes, since this will provide physical intuition into the effect of the isocurvature modes on the BAO.
\subsection{AD mode}
\noindent
The adiabatic mode is characterized by the requirement that the densities of all species are perturbed in proportion at some initial time such that
\begin{equation}
\delta_{c,i}=\delta_{b,i}=\frac{3}{4}\delta_{\gamma,i}=\frac{3}{4}\delta_{\nu,i},
\end{equation}
where the subscript $i$ labels the initial time. Or equivalently, using the relative entropy between two species $x$ and $y$ given by $\mathcal{S}_{xy}=\frac{\delta_x}{1+w_x}-\frac{\delta_y}{1+w_y}$, where $w_x$ and $w_y$ are the equation of state parameters of the species $x$ and $y$ respectively, we have that $\mathcal{S}_{xy}=0$ for all pairs of species at the initial time. In addition, all velocity divergences are initially unperturbed. Therefore, using the initial conditions for the adiabatic mode \cite{Bucher00}, the photon and baryon density contrasts are respectively given by 
\begin{align}
\nonumber
\delta_{\gamma}^{AD}&=\frac{\sqrt{3}}{k}e^{-k^2/k_D^2}\\
\nonumber
&\quad\times\int_0^{\tau} (1+R(\tau'))^{1/2}\sin{[kr_s(\tau)-kr_s(\tau')]}\\
\label{deltag_ad}
&\qquad\times F^{AD}(\tau')d\tau',\\
\nonumber
\delta_{b}^{AD}&=\frac{3}{4}\frac{\sqrt{3}}{k}e^{-k^2/k_D^2}\\
\nonumber
&\quad\times\int_0^{\tau} (1+R(\tau'))^{1/2}\sin{[kr_s(\tau)-kr_s(\tau')]}\\
\label{deltab_ad}
&\qquad\times F^{AD}(\tau')d\tau'.
\end{align}

\noindent
Thus, the adiabatic mode is only sourced by the gravitational driving term $F^{AD}$. This driving term can be approximated by $$F^{AD}(k,\tau)\approx 2k^2c_s^2j_{0}(kr_s)$$ on small and intermediate scales which reduces to $2k^2c_s^2$ at early times. On very large scales the above approximation breaks down, however, this does not affect our physical description of the BAO peak as these large-scale modes are well outside the horizon at decoupling and do not substantially influence the BAO features. The lack of an exact analytic expression for the driving term makes it difficult to derive exact analytic solutions for the time evolution of the photon and baryon density contrasts. Nevertheless, good approximations for the photon and baryon density contrasts are given by
\begin{equation}
\label{deltab_ad_approx}
\delta_{\gamma} =\frac{4}{3}\delta_{b}\approx 2kr_s {j}_1(kr_s) \times e^{-k^2/k_D^2}.
\end{equation}

Therefore, at early times $(kr_s(\tau)\ll 1)$ the density contrasts for the adiabatic mode, $\delta_{\gamma}\propto\delta_{b} \propto(1-\cos{kr_s})$ couple to a $\cos{kr_s}$ harmonic \cite{Hu_Sugiyama_1995}.\\

\begin{figure}[h]
\centering
\includegraphics[scale=.8]{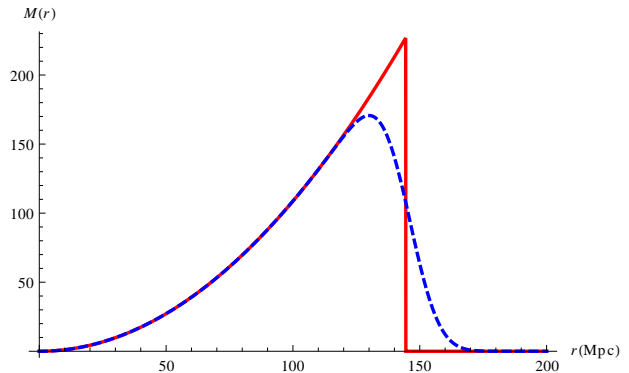}
\caption{Effect of Silk damping on the baryon mass profile for the AD mode at decoupling. The solid curve represents the baryon mass profile without the Silk damping correction, while the dashed curve represents the baryon mass profile with the Silk damping factor turned on. In the absence of the damping term, the peak is located at $r=r_s=144.5\mbox{ Mpc}$.}
\label{ad_damping}
\end{figure}

\begin{figure}[th!]
\centering
\subfigure[]{
\label{massprof_ad1}
{\includegraphics[scale=.95]{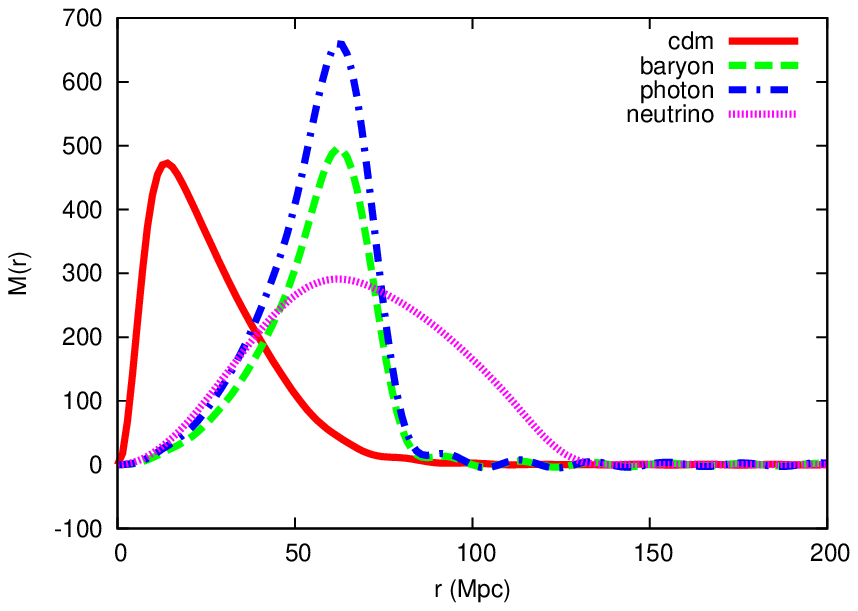}}}
\subfigure[]{
\label{massprof_ad2}
{\includegraphics[scale=.95]{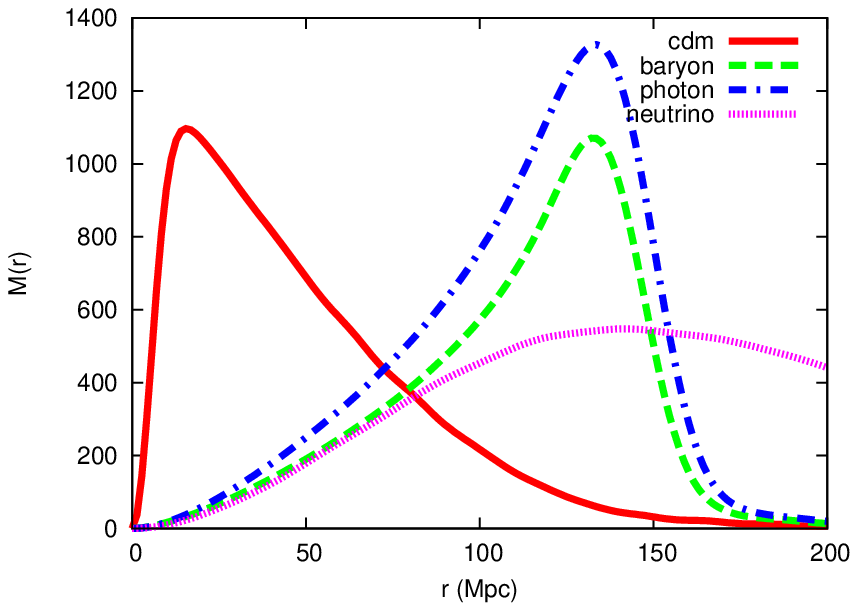}}}
\subfigure[]{
\label{massprof_ad4}
{\includegraphics[scale=.95]{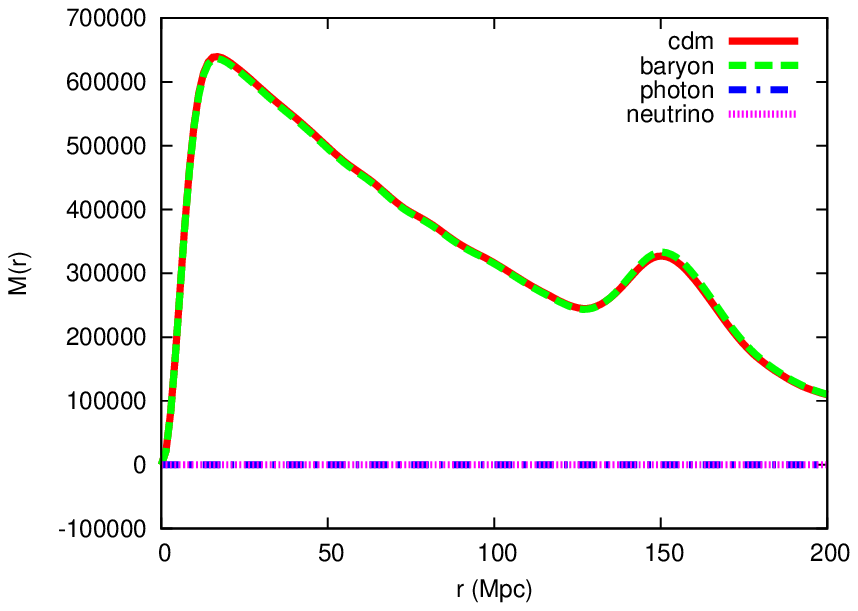}}}
\caption{Mass profile snapshots for the AD mode at different redshifts. The red, green, blue and purple curves respectively represent the CDM, the baryon, the photon and the neutrino mass profiles. \textbf{(a)} Well before decoupling ($z=3000$), \textbf{(b)} At decoupling ($z=1080$), \textbf{(c)} At late times ($z=0$). The units of the mass profile are arbitrary but correctly scaled between panels.}
\label{massprof_ad}
\end{figure}
\noindent
Now, in the perfect tight-coupling approximation, that is if we omit the Silk damping correction in the density contrast equations, the baryon mass profile is given by
\begin{align}
\label{anal_massp_ad}
M_b(r)\propto \left(1-\Heav{(r-r_s)}\right)\frac{\pi r^2}{2r_s^2}\propto\begin{cases}
r^2 \text{ for }r \leq r_s,\\
0 \text{ for }r > r_s,
\end{cases}
\end{align}
which is obtained by substituting the density contrast expression, without the Silk damping term, in the baryon mass profile expression. Here $H(x)$ is the Heaviside step function. We observe that in the absence of Silk damping, the baryon mass profile is quadratic at lower $r$ and sharply peaked at a distance $r(z)=r_s(z)$. This is illustrated in Figure \ref{ad_damping} where we show the effect of Silk damping on the baryon mass profile at decoupling. We see that when we include the Silk damping term, the BAO peak is smoothed, attenuated and shifted to lower $r$. We can understand these features as follows. As we approach decoupling, the coupling between photons and baryons weakens and the photon mean free path increases. Photons diffuse from overdensities to underdensities carrying baryons with them. Therefore, baryons leak out of the overdensity to both smaller and larger $r$, thereby smoothing and lowering the BAO peak. Due to the shape of the undamped mass profile (with no baryons on scales larger than the sound horizon), Silk damping has the effect of moving more baryons to larger scales. As a result, the BAO peak is at a slightly smaller distance than the sound horizon. As we will see later, Silk damping changes the shape of the mass profile for the adiabatic and isocurvature modes in different ways, due to the differing shapes of the undamped mass profiles. This has important consequences for our ability to use the BAO peak as a standard ruler.

\noindent
The redshift evolution of the mass profile for the AD mode has previously been studied in the literature  \cite{Eisenstein07}. Initially the overdensities of all species coincide. As time evolves, the photon pressure drives acoustic waves in the photon-baryon fluid, while neutrinos free stream at the speed of light and the CDM remains at its initial location. In Figure \ref{massprof_ad} we show the redshift evolution of the CDM, baryon, photon and neutrino mass profiles. Prior to decoupling, photons drag baryons at the sound speed, leaving behind a void of baryons. Thus, the initial baryon point-like overdensity evolves in a spherical shell while the CDM overdensity collapses at the origin under gravitational instability, and the neutrinos free stream.

\noindent
After decoupling, photons free stream while baryons, free from the photons, collapse into the CDM potential wells. The baryon overdensity continues to collapse, pulling matter from the surrounding underdense regions to the overdense regions. As the baryon velocity divergence does not decay instantaneously at decoupling \cite{Eisenstein07}, the baryons only stall later at $z\sim 500$ with the consequence that the BAO peak is closer to $150\mbox{ Mpc}$ than $140\mbox{ Mpc}$, the sound horizon size at decoupling. At $z=0$, the baryon mass profile displays two peaks, one near the origin and a second peak at approximately $150\mbox{ Mpc}$.

\subsection{NID mode}
\begin{figure}[th!]
\centering
\subfigure[]{
\label{massprof_nid1}
{\includegraphics[scale=0.95]{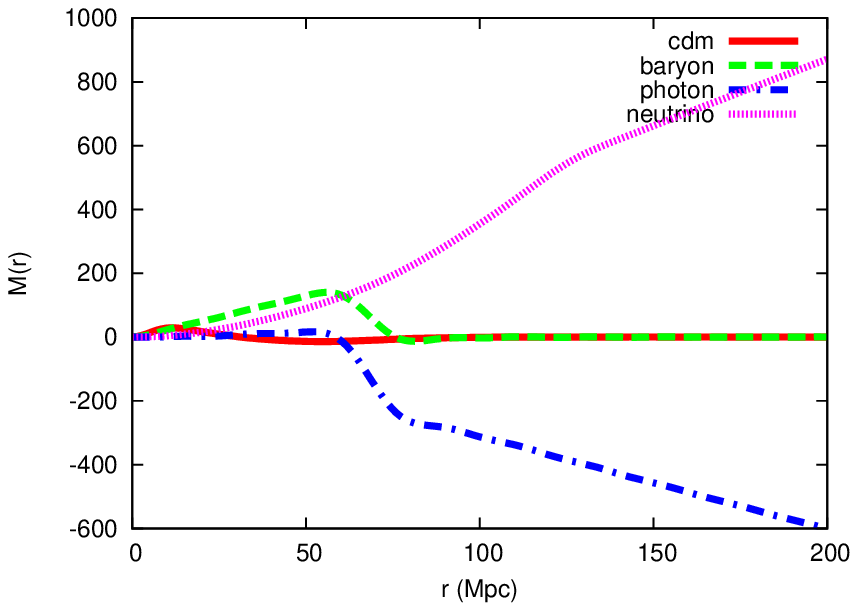}}}
\subfigure[]{
\label{massprof_nid2}
{\includegraphics[scale=0.95]{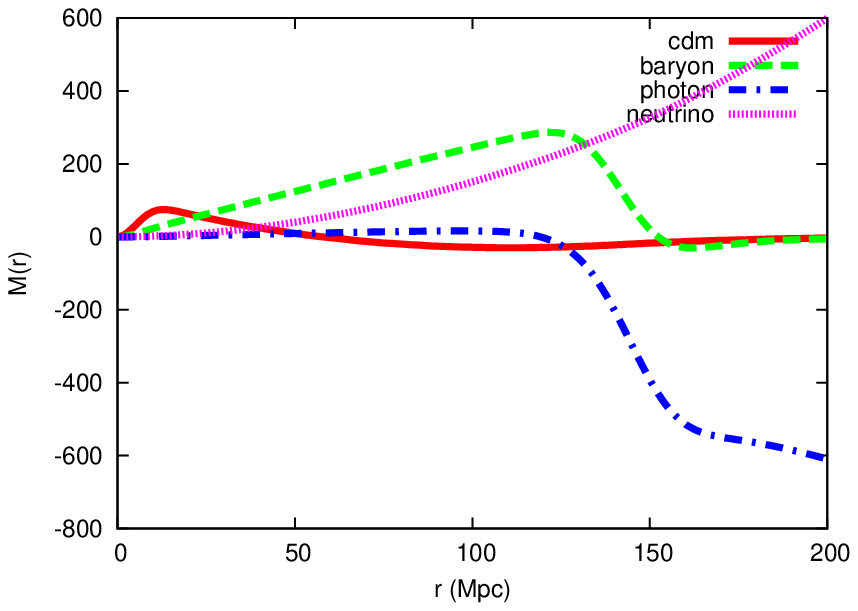}}}
\subfigure[]{
\label{massprof_nid4}
{\includegraphics[scale=0.95]{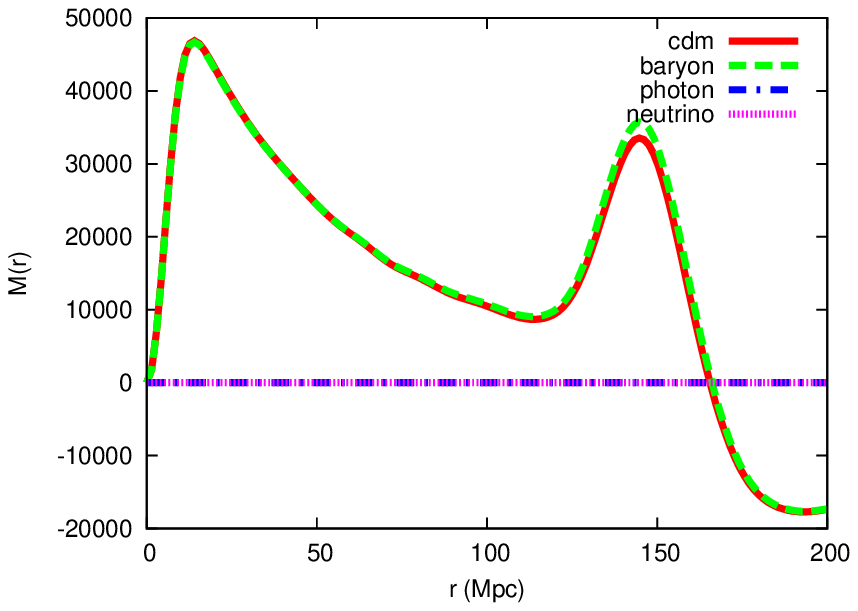}}}
\caption{Mass profile snapshots for the NID mode at different redshifts. The red, green, blue and purple curves respectively represent the CDM, the baryon, the photon and the neutrino mass profiles. \textbf{(a)} Well before decoupling ($z=3000$), \textbf{(b)} At decoupling ($z=1080$), \textbf{(c)} At late times ($z=0$). The units of the mass profile are arbitrary but correctly scaled between panels.}
\label{massprof_nid}
\end{figure}

\noindent
The NID mode arises when the densities of the matter components are initially unperturbed while the initial perturbation in the neutrino density is balanced by its photon counterpart, keeping the curvature unperturbed. The initial perturbations are as follows:
\begin{equation}
\delta_{c,i}=\delta_{b,i}=0,\quad\delta_{\gamma,i}=-\frac{R_{\nu}}{R_{\gamma}}\delta_{\nu,i}.
\end{equation}
These initial conditions imply that $A_S =0$, thus exciting the $\cos{kr_s}$ harmonic. The gravitational driving term contribution for this mode can be neglected without loss of accuracy, as the gravitational potential (related to $\dot{h}$), is initially unperturbed and only grows inside the horizon. This can also be understood by considering the right-hand side of equation (\ref{metric_field}). In the radiation dominated era, the photon and the neutrino density contrasts roughly cancel while the baryon and the CDM density contrasts remain small until the matter dominated era when they grow. The time evolution of the photon and the baryon density contrasts for the NID mode are given by
\begin{align}
\label{deltag_nid}
\delta_{\gamma}^{NID}&=-\frac{R_{\nu}}{R_{\gamma}}\sqrt{3}c_s\cos{k r_s}\times e^{-k^2/k_D^2},\\
\label{deltab_nid}
\delta_{b}^{NID}&=\frac{3}{4}\frac{R_{\nu}}{R_{\gamma}} \left(1-\sqrt{3}c_s\cos{k r_s}\right)\times e^{-k^2/k_D^2},
\end{align}
\noindent
where $R_{\nu}=\Omega_{\nu}/\Omega_{rad}$ and $R_{\gamma}=\Omega_{\gamma}/\Omega_{rad}$ are respectively the fractional energy densities of neutrinos and photons at early times. The pressure due to an initial localized photon overdensity creates a baryon underdensity that propagates due to its coupling to photons and perturbs, through gravitational interaction, the CDM (see Figure \ref{massprof_nid}). In addition, isocurvature perturbations grow once they enter the horizon. It follows that the BAO peak in the case of the NID mode has smaller amplitude than in the adiabatic case. With time, the baryon and the CDM overdensities grow by pulling more matter from their surroundings, thus creating underdense regions around them. Note that the mass profile of a given species can be negative since the species can be initially perturbed positively, corresponding to an overdensity or negatively, corresponding to an underdensity, with respect to the background level. The final baryon mass profile displays a deeper trough between the two peaks compared to the adiabatic case. Most importantly, though the baryon overdensity in the NID mode evolves at earlier times like the baryon overdensity in the adiabatic mode as they both excite $\cos{kr_s}$ harmonics, the final locations of the NID and the AD BAO peaks differ. At late times, the adiabatic mode becomes a superposition of sine and cosine waves, departing from the NID mode and with the undamped profile being convolved differently with Silk damping.

\begin{figure}[t]
\centering
\includegraphics[scale=.8]{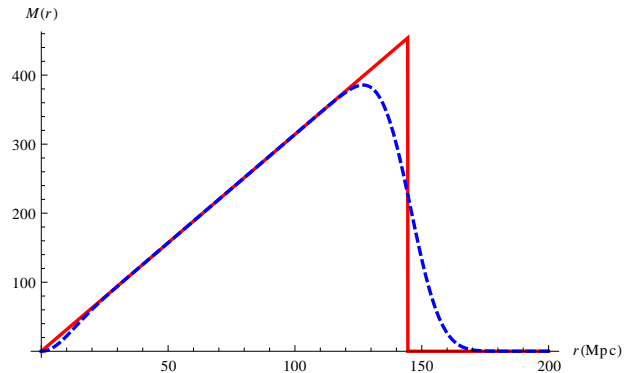}
\caption{Effect of Silk damping on the baryon mass profile for the NID mode at decoupling. In the absence of the damping term (solid curve), the peak is located at $r=r_s=144.5\mbox{ Mpc}$. The dashed curve takes into account the effect of Silk damping.}
\label{nid_damping}
\end{figure}

\noindent
Figure \ref{nid_damping} shows the effect of Silk damping on the baryon mass profile at decoupling for the NID mode. In the absence of Silk damping, the AD and the NID BAO peak locations would coincide. The undamped baryon mass profile for the NID mode is given by
\begin{align}
\label{anal_massp_nid}
M_b(r)\propto \left(1-\Heav{(r-r_s)}\right)r \propto\begin{cases}
r \text{ for }r \leq r_s,\\
0 \text{ for }r > r_s.
\end{cases}
\end{align}
As for the AD case, equation (\ref{anal_massp_nid}) is obtained by omitting the damping factor in equation (\ref{deltab_nid}) and substituting into equation (\ref{eq1}) for the mass profile. The baryon mass profile for the NID mode differs from the AD mode as it grows linearly with $r$ until $r(z)=r_s(z)$ then falls to zero. For this reason, the shift in the BAO peak location due to Silk damping is larger than in the case of the AD mode for which, as previously mentioned, the undamped mass profile is quadratic in $r$ for $r<r_s$. The difference in the shape of the undamped mass profile also sets the difference in the width of the BAO peak.

\subsection{NIV mode}
\noindent
Unlike the other isocurvature modes, the NIV mode, like the AD mode, shows no relative entropy perturbation in the density field at some initial time. All the density perturbations are zero initially. The main difference with the AD mode is in the velocity field where the neutrino velocity divergence starts perturbed, being compensated by the photon-baryon velocity. The initial perturbations are given by:
\begin{equation}
\theta_{c,i}=0,\quad \theta_{b,i}=\theta_{\gamma,i}=-\frac{R_{\nu}}{R_{\gamma}}\theta_{\nu,i}.
\end{equation}

\begin{figure}[ht!]
\centering
\subfigure[]{
\label{massprof_niv1}
{\includegraphics[scale=0.95]{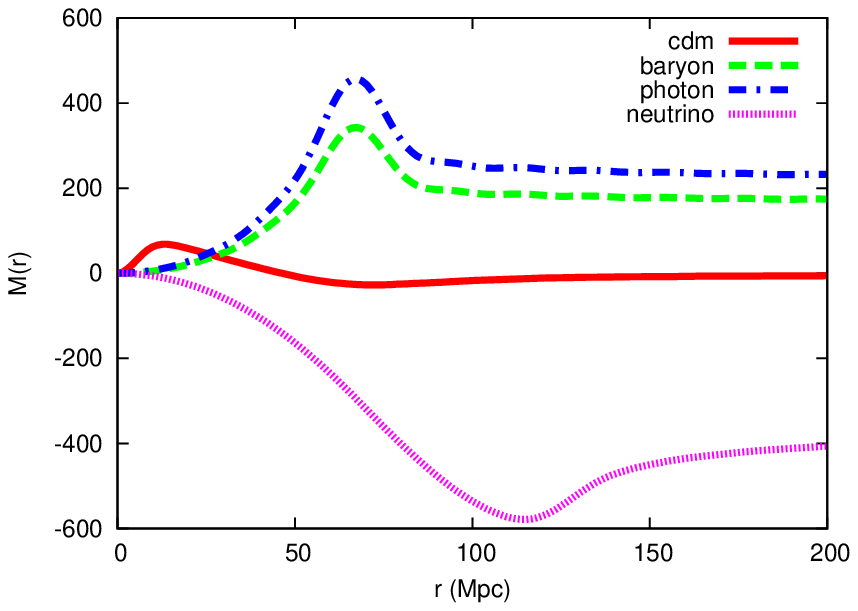}}}
\subfigure[]{
\label{massprof_niv2}
{\includegraphics[scale=0.95]{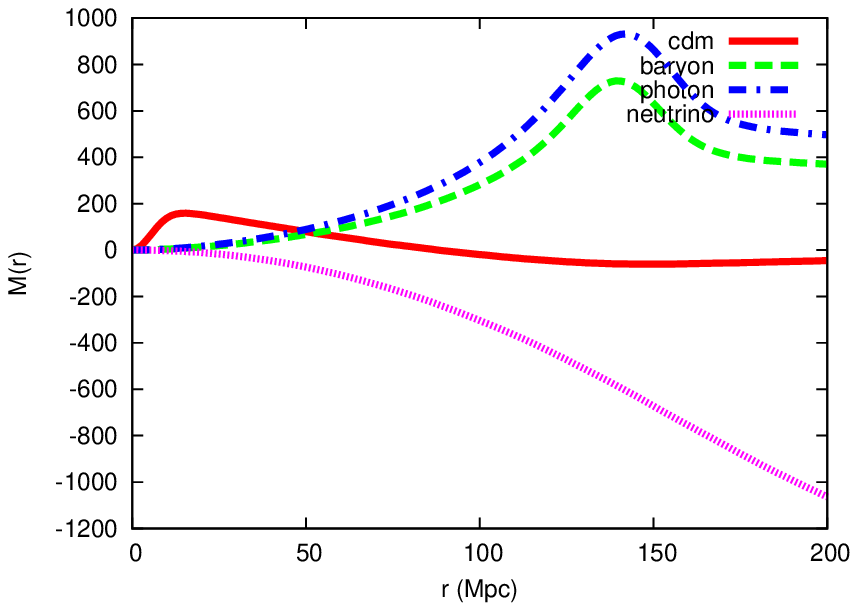}}}
\subfigure[]{
\label{massprof_niv4}
{\includegraphics[scale=0.95]{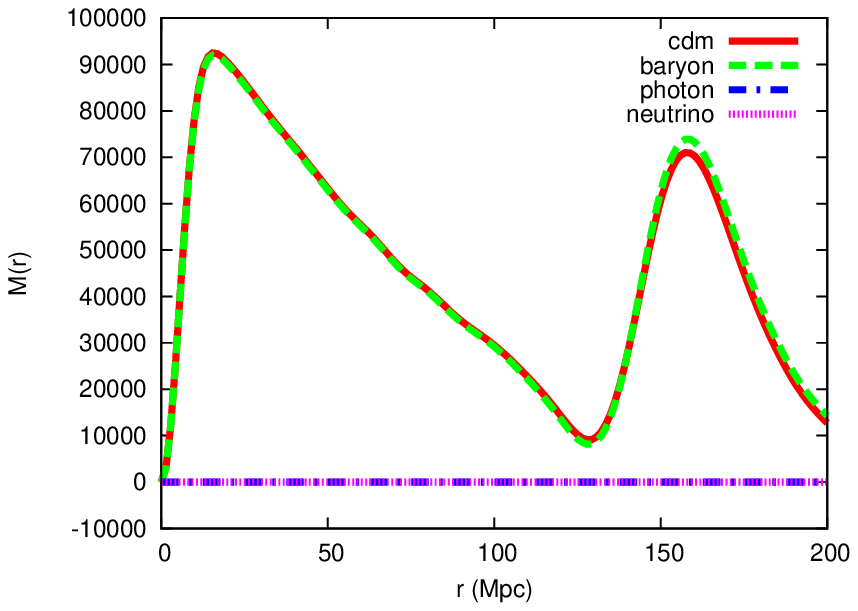}}}
\caption{Mass profile snapshots for the NIV mode at different redshifts. The red, green, blue and purple curves respectively represent the CDM, the baryon, the photon and the neutrino mass profiles. \textbf{(a)} Well before decoupling ($z=3000$), \textbf{(b)} At decoupling ($z=1080$), \textbf{(c)} At late times ($z=0$). The units of the mass profile are arbitrary but correctly scaled between panels.}
\label{massprof_niv}
\end{figure}

The NIV mode excites the $\sin{kr_s}$ harmonic, so that we can set $A_C=0$ in equation (\ref{deltag}). As in the case of the NID mode, the gravitational driving term contribution remains irrelevant at all times as all the densities start unperturbed and the perturbations only grow in the matter dominated era. The time evolution of the photon and baryon density contrasts for the NIV mode are given by

\begin{align}
\delta_{\gamma}^{NIV} =\frac{4}{3}\delta_{b}^{NIV}=\frac{R_{\nu}}{R_{\gamma}}\sqrt{3}\sin{k r_s(\tau)}\times e^{-k^2/k_D^2}.
\label{deltab_niv}
\end{align}
\noindent
The non-zero initial velocity divergence of baryons and photons pushes the baryons and photons from the origin, thus creating an overdensity at approximately the scale of the sound horizon and a plateau at larger scales, in the baryon and photon mass profiles. The redshift evolution of the baryon overdensity for the NIV mode is shown in Figure  \ref{massprof_niv}. This is similar to the NID case, except that the baryon mass profile remains positive at all times due to the initial plateau. The BAO peak is at a different location as the sine harmonic convolves differently with Silk damping, compared to the cosine harmonic.\\

\begin{figure}[b!]
\centering
\includegraphics[scale=.8]{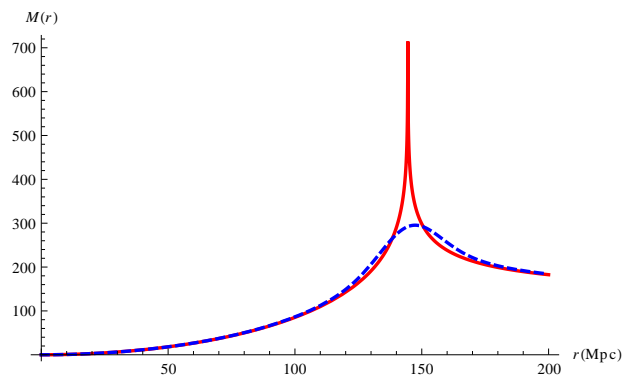}
\caption{Effect of Silk damping on the baryon mass profile for the NIV mode at decoupling. In the absence of the damping term (solid curve), the peak is located at $r=r_s=144.5\mbox{ Mpc}$. The dashed curve takes into account Silk damping.}
\label{niv_damping}
\end{figure}

\noindent
In the absence of the Silk damping correction, the undamped baryon mass profile for the NIV mode is given by
\begin{equation}
\label{anal_massp_niv}
M_b(r)\propto-\frac{r}{4}\ln{\frac{(r-r_s)^2}{(r+r_s)^2}}\quad\propto
\begin{cases}
r^2 \text{ for }r \ll r_s,\\
r^{-2} \text{ for }r \gg r_s.
\end{cases}
\end{equation}
The derivation of equation (\ref{anal_massp_niv}) is similar to the AD and NID cases. The undamped NIV mass profile grows quadratically with $r$ for $r<r_s$ and peaks at $r=r_s$ as for the AD case. However, the shift in the BAO peak location due to Silk damping is not as significant as it is for the AD and the NID cases for the simple fact that the undamped mass profile does not abruptly fall off to zero after the peak as in the previous cases but decreases as $r^{-2}$ before reaching a plateau of height proportional to $r_s$. This is due to the fact that the non-zero initial velocity of photons carries baryons beyond the sound horizon, compared to if they started from rest. Figure \ref{niv_damping} shows the effect of Silk damping on the undamped baryon mass profile. In contrast to the AD and NID cases, the BAO peak is slightly shifted to higher $r$. 

\subsection{CI \& BI modes}
\noindent
The CI and the BI modes have been well studied in the literature \cite{Bond_Efstathiou_1987,Chiba_Sugiyama_Suto_1994,Hu_Bunn_Sugiyama_1995}. The CI and BI modes are similar in that the perturbation starts in the CDM density contrast and the baryon density contrast respectively while the other species are initially unperturbed. This can be written at some initial time as
\begin{equation}
\delta_{c,i}=1,\quad\delta_{b,i}=\delta_{\gamma,i}=\delta_{\nu,i}=0,
\end{equation}
for the CI mode, and as
\begin{equation}
\delta_{c,i}=0,\quad\delta_{b,i}=1,\quad\delta_{\gamma,i}=\delta_{\nu,i}=0,
\end{equation}
for the BI mode. The CI and BI initial conditions dictate that $A_S=-\frac{8}{\sqrt{3}k}\Omega_{c,0}$ for the CI mode and $A_S=-\frac{8}{\sqrt{3}k}\Omega_{b,0}$ for the BI mode, while $A_c=0$ in both cases, thus exciting the $\sin{kr_s}$ harmonic \cite{Kodama86, Bucher00}. The constants $\Omega_{c,0}$ and $\Omega_{b,0}$ are respectively the CDM and the baryon densities today. The driving term is negligible in the radiation domination era as the photon and the neutrino densities are initially unperturbed but becomes important in the matter domination era as the matter perturbation sources the gravitational potential \cite{Hu_Sugiyama_1994}.

\begin{figure}[t!]
\centering
\subfigure[]{
\label{massprof_ci1}
{\includegraphics[scale=0.95]{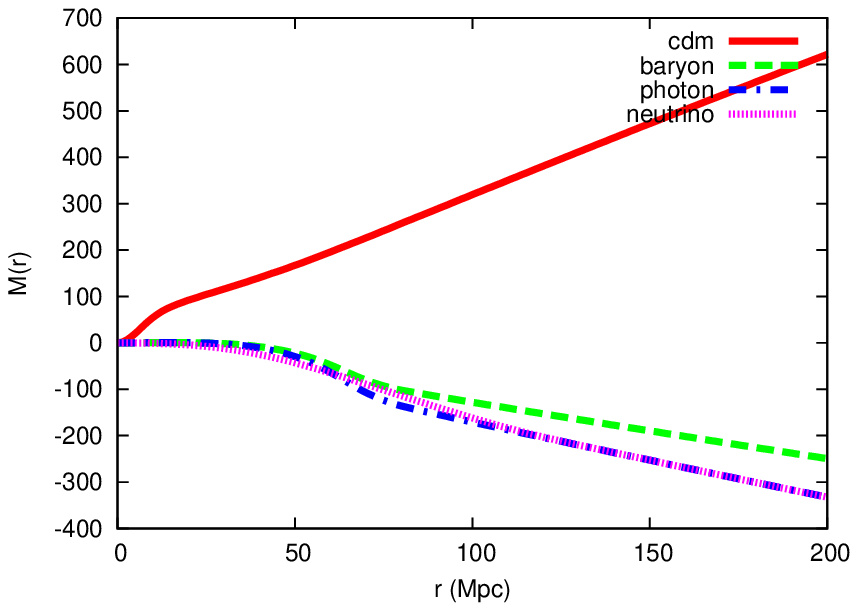}}}
\subfigure[]{
\label{massprof_ci2}
{\includegraphics[scale=0.95]{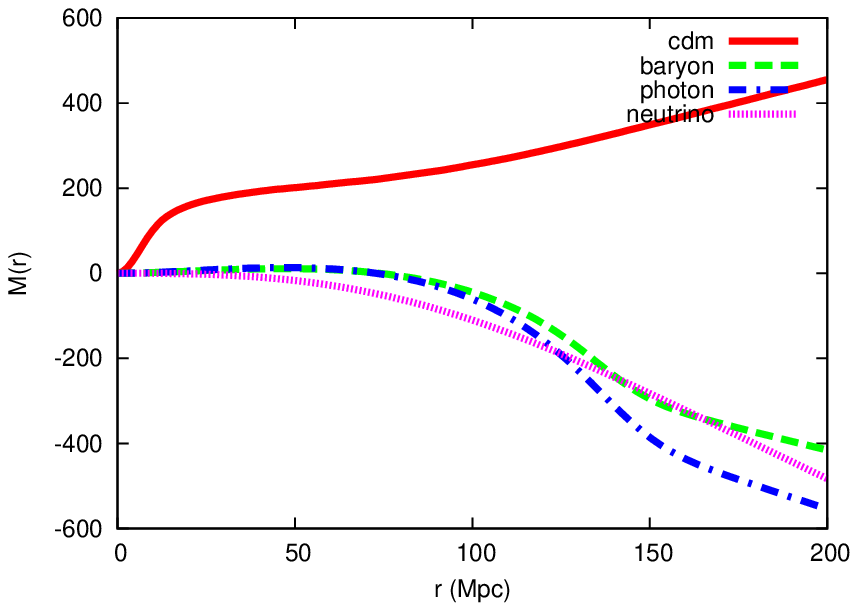}}}
\subfigure[]{
\label{massprof_ci4}
{\includegraphics[scale=0.95]{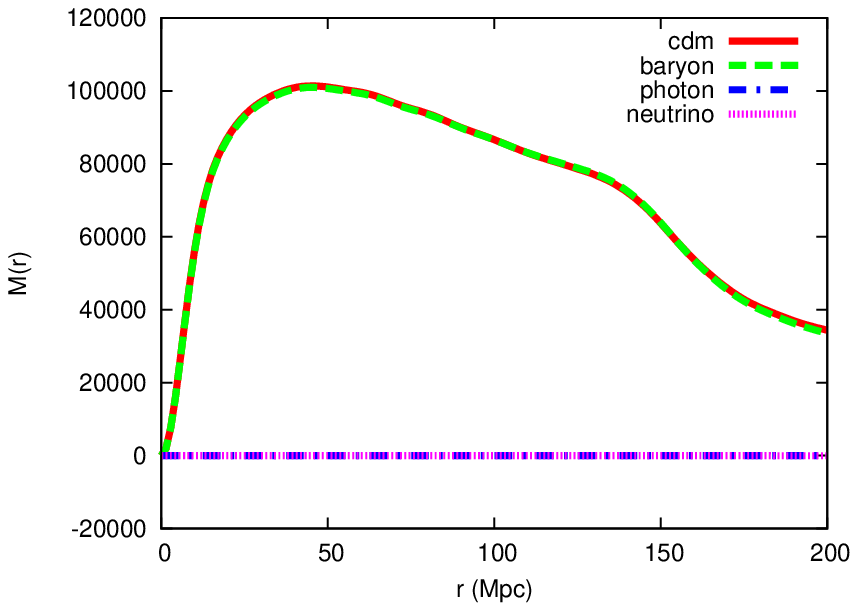}}}
\caption{Mass profile snapshots for the CI mode at different redshifts. The red, green, blue and purple curves respectively represent the CDM, the baryon, the photon and the neutrino mass profiles. \textbf{(a)} Well before decoupling ($z=3000$), \textbf{(b)} At decoupling ($z=1080$), \textbf{(c)} At late times ($z=0$). The units of the mass profile are arbitrary but correctly scaled between panels.}
\label{massprof_ci}
\end{figure}

The time evolution of the photon and baryon density contrasts for the CI and BI modes is given by \cite{Hu_Sugiyama_1994}
\begin{align}
\nonumber
\delta_{\gamma}^{CI} &=-\frac{8}{3}\Omega_{c,0}\frac{\sqrt{3}}{k}\sin{kr_s(\tau)}\times e^{-k^2/k_D^2}\\
\nonumber
&\quad+\frac{\sqrt{3}}{k}\int_0^{\tau} (1+R(\tau'))^{1/2}\sin{[kr_s(\tau)-kr_s(\tau')]}\\
\label{deltag_ci}
&\qquad\times F^{CI}(\tau')d\tau',\\
\nonumber
\delta_{b}^{CI} &=-2\Omega_{c,0}\frac{\sqrt{3}}{k}\sin{kr_s(\tau)}\\
\nonumber
&\quad+\frac{3}{4}\frac{\sqrt{3}}{k}\int_0^{\tau} (1+R(\tau'))^{1/2}\sin{[kr_s(\tau)-kr_s(\tau')]}\\
\label{deltab_ci}
&\qquad\times F^{CI}(\tau')d\tau'\times e^{-k^2/k_D^2},
\end{align}
for the CI mode, and by
\begin{align}
\nonumber
\delta_{\gamma}^{BI} &=-\frac{8}{3}\Omega_{b,0}\frac{\sqrt{3}}{k}\sin{kr_s(\tau)}\\
\nonumber
&\quad+\frac{\sqrt{3}}{k}\int_0^{\tau}(1+R(\tau'))^{1/2}\sin{[kr_s(\tau)-kr_s(\tau')]}\\
\label{deltag_bi}
&\qquad\times F^{BI}(\tau')d\tau'\times e^{-k^2/k_D^2},\\
\nonumber
\delta_{b}^{BI} &=1-2\Omega_{b,0}\frac{\sqrt{3}}{k}\sin{kr_s(\tau)}\\
\nonumber
&\quad+\frac{3}{4}\frac{\sqrt{3}}{k}\int_0^{\tau} (1+R(\tau'))^{1/2}\sin{[kr_s(\tau)-kr_s(\tau')]}\\
\label{deltab_bi}
&\qquad\times F^{BI}(\tau')d\tau'\times e^{-k^2/k_D^2},
\end{align}
for the BI mode. Equations (\ref{deltag_ci}-\ref{deltab_bi}) are exact but require a perfect knowledge of the gravitational driving term. This makes the derivation of simple explicit analytic expressions for the CI and BI modes harder as compared to the AD, NID and NIV modes. Therefore, we do not discuss the effect of Silk damping on the BAO peak feature for these modes. However, one thing to notice is the $k^{-1}$ dependence of the baryon density contrast for the CI and BI modes that washes out perturbations on small scales while amplifying them on large scales. This redistribution of power results in a flattening of the baryon mass profile for these modes. On small scales, the $k^{-1}$ and the Silk damping factors have similar effects on the BAO peak as they both suppress perturbations on these scales. However there are two main differences. Firstly, Silk damping does not act on large scales while the $k^{-1}$ factor amplifies large scale perturbations. Secondly, Silk damping only becomes significant around recombination while the $k^{-1}$ factor redistributes the power at all times, hindering the development of a well defined BAO peak but producing a knee instead.

\noindent

For the CI mode, an overdensity in the CDM component tends to affect, through gravitational attraction, the baryon density component by gathering baryons into an overdensity but the photon pressure opposes this process until decoupling. 
\begin{figure}[ht!]
\centering
\subfigure[]{
\label{massprof_bi1}
{\includegraphics[scale=0.95]{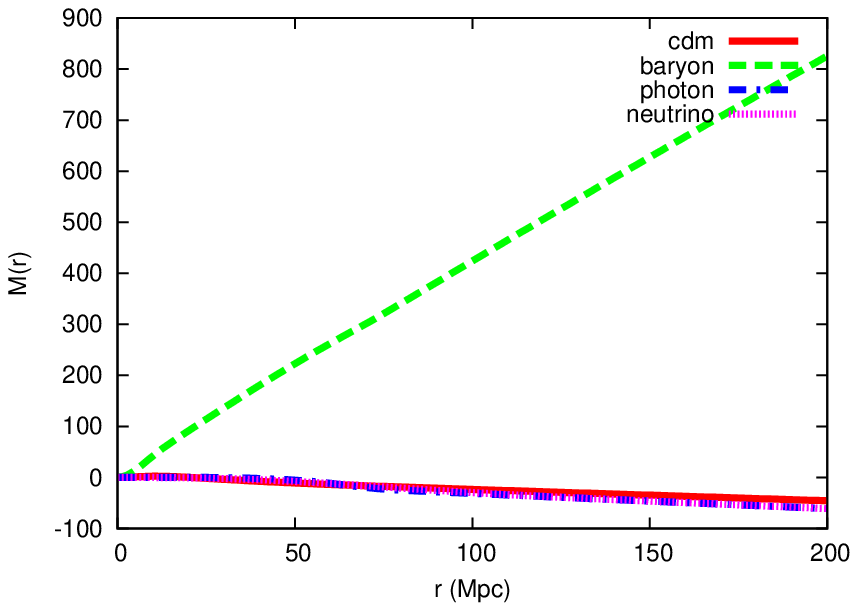}}}
\subfigure[]{
\label{massprof_bi2}
{\includegraphics[scale=0.95]{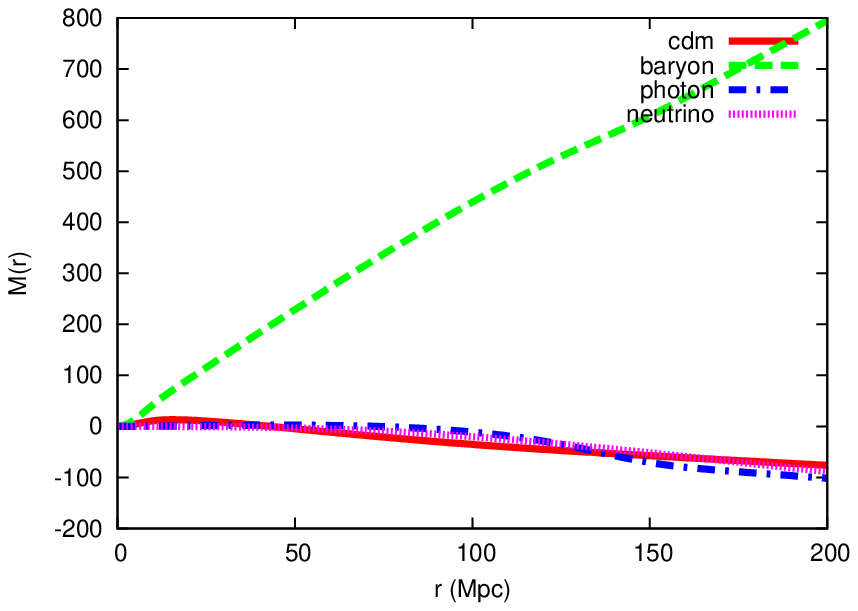}}}
\subfigure[]{
\label{massprof_bi4}
{\includegraphics[scale=0.95]{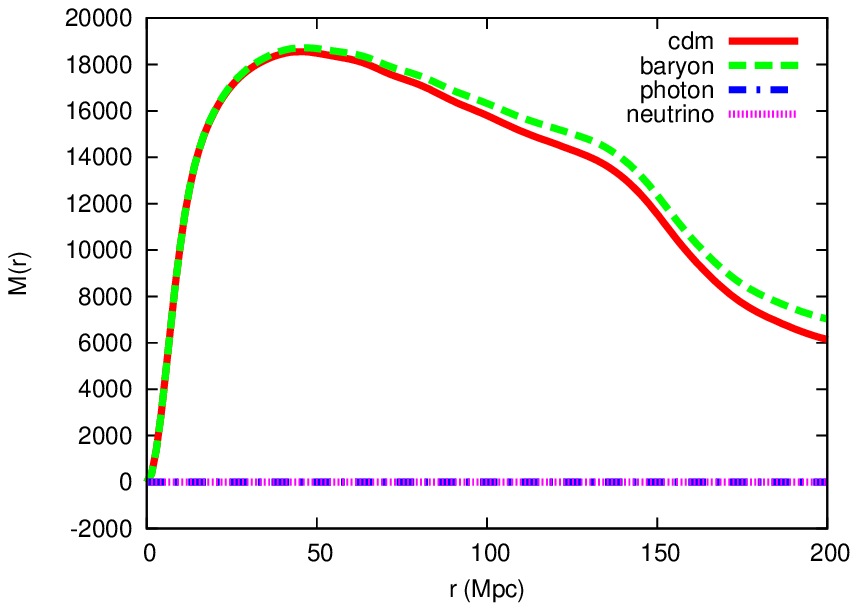}}}
\caption{Mass profile snapshots for the BI mode at different redshifts. The red, green, blue and purple curves respectively represent the CDM, the baryon, the photon and the neutrino mass profiles. \textbf{(a)} Well before decoupling ($z=3000$), \textbf{(b)} At decoupling ($z=1080$), \textbf{(c)} At late times ($z=0$). The units of the mass profile are arbitrary but correctly scaled between panels.}
\label{massprof_bi}
\end{figure}

One should note that an initial overdensity in the photon component would more easily affect the baryon component than an initial overdensity in the CDM component, the reason being the high photon pressure at earlier times. Therefore, the perturbation takes longer to imprint ripples onto the homogeneous sea of baryons. Figure \ref{massprof_ci} represents the time evolution of the baryon mass profile for the CI mode. Prior to decoupling, the CDM overdensity grows but does not significantly affect the baryon component. After decoupling a baryon overdensity develops through gravitational interaction with the CDM but fails to display a well defined BAO peak.\\

For the BI mode, an initial overdensity in the baryon component affects the CDM component through gravitational attraction, but does not significantly grow due to the photon pressure at earlier times that tends to widen and even wash out the baryon overdensity as can be seen in Figure \ref{massprof_bi}. With a similar process as for the CI mode, the overdensity becomes a knee at late times.%
\noindent
Although the Silk damping affects the CI and BI modes, its effect is not as significant as in previous cases (for a discussion of this see \cite{Hu_Bunn_Sugiyama_1995,Rich2001}). We recall that the Silk damping tends to suppress power on small scales while these modes are already significantly reduced by the  $k^{-1}$ factor for the CI and BI modes. In addition, the fact that the CI and BI modes fail to display a well defined BAO peak makes less noticeable the effect of Silk damping on the BAO peak.\\

\subsection{Time evolution of the BAO peak position}

We saw in previous subsections that in the absence of Silk damping, the BAO peak location for all the modes would coincide at all times as the acoustic wave in the photon-baryon fluid propagates at the same sound speed irrespective of the initial conditions. Here we consider the effect of Silk damping on the evolution of the BAO peak location for different modes.

\begin{figure}[H]
\centering
\includegraphics[scale=.6]{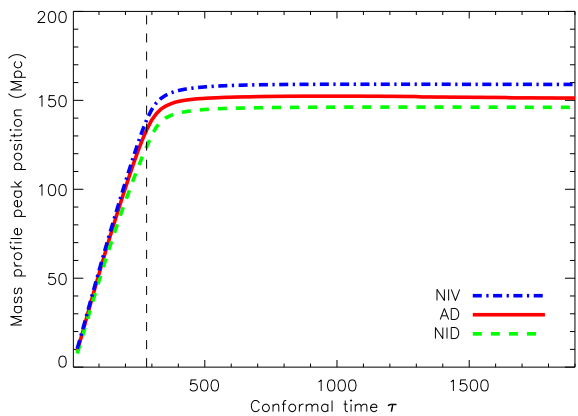}
\caption{Time-evolution of the baryon mass profile peak location (see Figures \ref{massprof_ad}, \ref{massprof_nid} and \ref{massprof_niv}) for the AD, NID and NIV modes. These curves were obtained numerically from the evolution of the mass profile curves. The dashed vertical line indicates the epoch of recombination.}
\label{BAOpeak_compare}
\end{figure}

Figure \ref{BAOpeak_compare} shows the time evolution of the BAO peak position for the AD, NID and NIV modes. We do not include the CI and BI modes as they fail to display a defined BAO peak. At early times, the BAO peak positions for the AD, NID and NIV coincide as the Silk damping factor equals one at early times. With time, the damping scale $k_D^{-1}$ increases as the photon-baryon coupling weakens, the three modes depart from each other and the separation increases up to decoupling. At decoupling, $k_D\approx 0.15\mbox{ Mpc}^{-1}$, leading to a separation of about $15\mbox{ Mpc}$ between the NID and NIV BAO peak positions. After decoupling, though the BAO peak position still increases until $z\approx 500$ due to the bulk velocity, the separation between the modes remains constant until today.

\section{Impact of isocurvature modes on dark energy constraints}
\label{sec:FM}

The aim of this section is to quantify the potential impact of isocurvature modes on dark energy studies based on current- \& next-generation datasets.
\subsection{Statistical Formalism}
A convenient way of quantifying the accuracy with which cosmological parameters can be measured from a given dataset is the Fisher matrix formalism (see \cite{TTH} for a review). If \textbf{x} is our observable (the CMB or the galaxy power spectrum in our case), it can be modeled as a N-dimensional random variable whose probability distribution $\l(\bf{x}; \bftheta)$ depends on a vector of cosmological parameters $\bftheta$ that we wish to estimate. $\l(\bf{x}; \bftheta)$ is also known as the likelihood of observing a set of data given a model characterized by $\bftheta$.\\

In this study, we consider a spatially flat cosmological model described by the following parameters: the baryon density $\omega_b$, the CDM density $\omega_c$, the density of the dark energy component $\Omega_X$, the optical depth $\tau$, the spectral index $n_s$ and the scalar amplitude $A_s$. 
We allow for dark energy models that vary with time and parametrize the dark energy equation of state as $w(a) = w_0 + (1-a) w_a$ \cite{Chevallier:2000qy,linder-2003-90} where $a=1/(1+z)$ and $w_0$ and $w_a$ are included in the parameter space. For the isocurvature modes, we adopt the parametrization implemented in \cite{Moodley04}, where the AD, CI, NID, NIV modes and their cross-correlations are described by 10 parameters, $z_{ij}$, measuring the fractional contributions of the various correlations (auto and cross) to the total power spectrum. We do not consider the BI mode as it has the same spectra as the CI mode. In terms of these fractional parameters, the total isocurvature fraction $f_{\ISO}$ is given by
\be
f_{\ISO}=\frac{z_{\ISO}}{z_{\ISO}+z_{\adE}},
\ee
where $z_{\ISO}=\sqrt{1-z^{2}_{\adE}}$ is the total isocurvature contribution.

Defining the auto- and cross-correlated primordial power spectra as follows
\be
P_{ij}(k) = A_{ij} k^{n_{ij}-1},
\ee

and the observed full power spectrum that includes adiabatic and all different modes as
\be
P(k)=\sum_{i,j,i\le j}A_{i,j} P_{i,j}(k),
\ee

the spectral indices of the cross correlated modes are given by $n_{ij} = \frac{n_{ii}+n_{jj}}{2}$ with their amplitudes $A_{ij} \propto z_{ij}$.  The constraint $\sum_{i,j=1}^{10} z^2_{ij}  = 1$ requires that the isocurvature parameters $z_{ij}$ exist on the surface of a $9$ dimensional sphere of unit radius. For further details on how this parametrization relates to others in the literature, see \cite{Bean06}. The parameter $A_s$ rescales the unit power CMB temperature spectrum to its usual amplitude as, $C_{\ell}=13000\mu\mbox{K}^2\mbox{ A}_s\mbox{ }\hat{C}_{\ell}$, where $\hat{C}_{\ell}$ is the fiducial CMB temperature spectrum with unit power.

\begin{table}[htb]
\begin{center}
\begin{tabular}{cccccccc}
\hline
\hline
&&Fiducial & model\\
\hline
$\omega_b$& $\omega_c$& $\Omega_X$& $\tau$& $n_s$&  $A_s$& $w_0$&$w_a$\\
\hline
0.02205& 0.12495& 0.7& 0.1& 1.0& 15.7& -1.0& 0.0\\
\hline
\hline
\end{tabular}
\end{center}
\caption{Values for the parameters of the fiducial cosmological model.}
\label{fid_values}
\end{table}

The Fisher matrix is defined by
\be
F_{ij} =  - \left\langle \frac{\partial^2 \ln \l(\bf{x}; \bftheta)}{\partial \theta_i \partial \theta_j}\right\rangle.
\label{eq:FM} 
\ee The Cramer-Rao inequality shows that $F^{-1}_{ii}$ is the smallest variance possible for an 
unbiased estimator of the parameter $\theta_i$. In this case, $F^{-1}$ is the most optimistic covariance matrix of the dataset 
\cite{TTH} and the forecasted error bar for ${\theta}_i$ is 
\be
\sigma_i = \sqrt{\bra{F^{-1}}_{ii}}.
\ee
The fiducial model $\bftheta$ around which the Fisher matrix is computed 
is chosen to be a $\Lambda$CDM universe with adiabatic initial conditions. The cosmological parameter values for the fiducial model are given in table \ref{fid_values}. We use the CMB forecasts from the {\planck} experiment in addition to each LSS data set and compute the full Fisher matrix $
F_{ij} = F^{CMB}_{ij} + F^{LSS}_{ij}$ for the cosmological parameter set.

\subsubsection{Large Scale Structure (LSS) surveys}

Over the next decade, the increase in the number and quality of data from LSS surveys will drive fundamental improvements in precision cosmology. As these galaxy surveys cover increasingly larger 
volumes, they will provide unprecedented probes of scales at which significant cosmological information is available.

The potential of the BAO method as a powerful source of cosmological information has been recognized and measuring the BAO peak at multiple redshifts is now regarded as the primary science of major future LSS surveys. We consider two such BAO experiments, one of which is
the Baryon Oscillation Spectroscopy Survey ({\BOSS}).  {\BOSS} will measure the redshifts of 1.5 million luminous red galaxies (LRGs) over a quarter of the sky to a depth of $z=0.7$. In addition to being a redshift survey,  {\BOSS} will be the first attempt to resolve the BAO peak in the high-z density field $(2<z<3)$, as allowed by 
mapping absorption lines from neutral hydrogen, in the spectra of 160 000 distant quasars \cite{BOSS_white_paper}. 

The Advanced Dark Energy Physics Telescope ({\ADEPT}) is the second future LSS survey that we consider. It is a space-based experiment aiming at 
mapping galaxies in the redshift range $1<z<2$ and over $28,600$ sq. deg. of the sky \cite{ADEPT}. 

The BAO peak manifests as oscillations in the matter power spectrum with the size of the sound horizon determining the frequency of these oscillations. However, the matter power spectrum is a rich statistic whose features at different scales provide specific cosmological information.
The matter power spectrum is defined as
\be
P(k,z) =   D(z)^2 P_{prim}(k)T^2(k) 
\ee where $D(z)$ is the growth rate of structure, $P_{prim}(k)$ is the primordial power spectrum and $T(k)$ is the transfer function. 
The first source of information is the baryon acoustic oscillations, with their wavenumber $k = 2\pi/r_s$ being set by the size of the sound horizon at decoupling $r_s$. Since this characteristic scale is calibrated by the CMB, measuring the wavelength of these oscillations both in the radial and tangential directions delivers $D_A(z)$ and $H(z)$ respectively. The overall shape of the matter power spectrum is a second source of information.  Any features which deviate from a power law, such as the turnover, provides an additional characteristic scale which is required by the Alcock-Paczynski test to be isotropic \cite{AP}. Lastly, the overall time evolution of the amplitude informs us about $D(z)$, the growth rate of structure. 

In reality we measure the power spectrum as mapped by galaxies which are biased tracers of the 
underlying matter distribution. We can write the galaxy power spectrum as $P_g(k,z) = b(k,z)^2 P(k,z)$ where $b(z,k)$ represents this bias resulting from the effects of galaxy formation and evolution. On the scales of the BAO, the bias can be regarded as smooth, i.e., $b(z,k) =b(z)$. Any scale dependence that is not taken into account is not likely to lead to oscillations in Fourier space \cite{SE05}. 
Furthermore, the galaxy power spectrum measured in redshift space is distorted relative to the 
power spectrum in real space as a result of galaxy peculiar velocities. Because galaxies moving towards an overdensity along the line of sight appear further away than equidistant galaxies moving in the tangential direction, structures appear "squeezed" in redshift space, with the amount of the distortion determined by the growth rate. On large scales this has been shown to give rise to an angle-dependent distortion which 
leads to a multiplicative change in the power that is a function of angle, ie. 
$P_{g,\beta} = \bra{1+ \beta(z) \mu^2}^2 P_g(k,z)$ where $\mu$ is 
the angle with respect to the line of sight and $\beta = f/b$ where
\be
f = \frac{\partial \text{ln} D(a)}{\partial \text{ln} a} \simeq \Omega_m(z)^{0.6}.
\ee
We follow \cite{SE07} and include a free shot noise parameter as well as a redshift distortion parameter $\beta(z)$ in each bin. 

Assuming the likelihood function of the band powers of the galaxy power spectrum to be Gaussian, the Fisher 
matrix can be approximated as \cite{TTH,SE03}:
\ber
F^{LSS}_{ij} &=& \int^{\vec{k}max}_{\vec{k}min} \frac{\partial \ln P(\vec{k})}{\partial p_i} \frac{\partial \ln P(\vec{k})}{\partial p_j} V_{eff}(\vec{k}) \frac{d\vec{k}}{2\bra{2\pi}^3} \nonumber\\
&=& \int^{1}_{-1} \int^{kmax}_{kmin} \frac{\partial \ln P(k,\mu)}{\partial p_i} \frac{\partial \ln P(k,\mu)}{\partial p_j} \nonumber\\
&&\times V_{eff}(k,\mu) \frac{2\pi k^2 dk d\mu}{2\bra{2\pi}^3}
\label{Fij}
\eer
where, 
\be
V_{eff}(k,\mu) = \left[ \frac{\bar{n}_g P_g(k)(1 + {\beta}{\mu}^2)^2}{\bar{n}_g P_g(k)(1 + {\beta}{\mu}^2)^2 + 1}  \right]^2V,
\ee
$\vec{r}$ is the unit vector along the line of sight and $\vec{k}$ is the wave vector with norm $k = |\vec{k}|$. Here $V$ is the survey volume contained in a given redshift bin and $\bar{n}_g(\vec{r})$ is the selection function of the survey, dictating the a priori expectation value for the comoving number density of galaxies. We take this to 
be a constant.  $V_{eff}$ is the effective volume of the survey and takes into account the 
impact of the shot noise from undersampled regions \cite{EHT}.

The derivatives of the power spectrum with respect to the cosmological parameters in table \ref{fid_values} and to the isocurvature parameters are respectively shown in Figures \ref{fig:derivs} and \ref{fig:derivs2} in the appendix. 

The {\BOSS} and {\ADEPT} survey parameters are summarized in Table \ref{table_specs}.  
Note that the value $k_{min}$ is always taken as the lowest possible and has been shown to have a 
negligible effect on the error forecasts. The smallest scale included, given by $k_{max}$, in the analysis does 
however impact on the results \cite{BAO_DE_Constraints, Rassat}. Following \cite{Rassat} we adopt conservative values for $k_{max}$ by 
requiring $\sigma(R)=0.2$ at a corresponding $R=\frac{\pi}{2k}$ where $\sigma(R)$ is defined 
similarly to the normalization  $\sigma_8 \equiv \sigma(R = 8h^{-1}\mbox{Mpc})$, but for a general scale R. For the {\BOSS} $z=3$ bin, we restrict $k_{max} = 0.3\mbox{ h Mpc}^{-1}$.  Although the application of the Fisher matrix formalism to the case of a Lyman-$\alpha$ forest is different to that of a galaxy survey in various subtle ways, the measurement of the BAO scale from a Lyman-$\alpha$ survey has however been explored extensively in \cite{McDonald}. Assuming typical values, we follow their prescription to estimate the shot noise contribution (1/$n_g$)for the $z=3$ bin of the {\BOSS} survey.  Note that $b$ for the $z=3$ bin is our estimate the linear bias between the flux and density field.

\begin{table}[htb]
\begin{center}
\begin{tabular}{llllll}
\hline
\hline
  & & {\BOSS} & & &  \\
\hline
n$_g$ & z & k$_{max}/ \mbox{h Mpc}^{-1}$ & b & V/ Gpc$^3$ & Area/deg$^2$ \\
	\hline
 $3\times10^{-4}$ & $z<0.35$ & 0.09 & 2.13 & 0.74 & 10,000 \\
$3\times10^{-4}$  & $0.35<z<0.6$ & 0.11 & 1.25 & 2.83 & 10,000 \\
 $3.1\times10^{-2}$  & $2 < z< 3$ & 0.3 & -0.18 & 2.48 & 6000 \\
\hline 
\hline
 & & {\ADEPT} & & & \\
\hline
n$_g$ & z & k$_{max}/ \mbox{h Mpc}^{-1}$ & b & V/ Gpc$^3$ & Area/deg$^2$ \\
	\hline
$3\times10^{-4}$ & $1<z<1.25$ & 0.16 & 2.97 & 17.7  & 28600 \\
$3\times10^{-4}$  & $1.25 < z< 1.5$ & 0.11 & 3.21 & 19.7  & 28600 \\
 $3\times10^{-4}$  & $1.5 < z< 1.75$ & 0.22 &  3.44 &  21.0  & 28600 \\
  $3\times10^{-4}$  & $1.75 < z< 2$ & 0.25 & 3.67 & 21.7  & 28600 \\
\hline
\hline
\end{tabular}
\end{center}
\caption{Table summarizing the survey parameters for {\BOSS} and {\ADEPT}, 
for different redshift bins (centered at the middle of the redshift bin).}
\label{table_specs}
\end{table}

\subsubsection{Cosmic microwave background (CMB) surveys}
The CMB data primarily provides information about the initial conditions of our Universe in this analysis. Non-adiabatic initial conditions lead to very distinct features in the temperature anisotropies, with
isocurvature modes producing acoustic oscillations that are out of phase with the adiabatic mode and hence a set of peaks in the temperature anisotropy power spectrum that are slightly shifted. 
Furthermore, CMB polarization provides a robust signature of isocurvature perturbations \cite{Bucher01}.  The latest WMAP data has confirmed that the initial perturbations were mainly of adiabatic type \cite{Komatsu10} with the possible presence of a subdominant isocurvature contribution, which could be detected in future high-precision experiments such as {\planck} \cite{Langlois04}. The higher resolution of {\planck} over WMAP will allow for the measurement of the CMB power spectrum on much smaller scales and the use of 9 observational bands will improve the modeling of astrophysical foregrounds. 

We follow the analysis in \cite{DETF} and model the {\planck} dataset as CMB 
temperature and polarization maps of 80$\%$ of the sky measured in the two 
frequency bands where the CMB signal dominates.  The details of the 
experiment are given in Table \ref{planck}. The maps are taken to have no 
foreground contribution, assuming that the other frequency channels 
can be used to remove them. The remaining 20$\%$ of the sky is assumed to be 
contaminated by galactic emission. We exclude polarization data at $\ell<30$ in order to weaken the forecasted constraint on the optical depth to $\sigma(\tau)=0.01$ in agreement with studies that include foreground modeling \cite{T01}.

\begin{table}[htb]
\begin{center}
\begin{tabular}{llllll}
\hline
\hline
 $\ell_{max}^T$ &  $\ell_{max}^P$ & $\nu /$ GHz &  $\theta_b$ & $\Delta_T$($\mu$K) & $\Delta_P$($\mu$K)\\
  \hline 
  2000 & 2500 & 143 & 8' & 5.2 & 10.8 \\
  &  & 217 & 5.5' & 11.7 & 24.3 \\
\hline
\hline
\end{tabular}
\end{center}
\caption{Summary of the experiment specifications for {\planck}.}
\label{planck}
\end{table}

For the CMB, the Fisher matrix is computed using
\be
F^{CMB}_{ij} = \sum_{\ell} \sum_{X,Y} \frac{\partial C_{X \ell}}{\partial p_i}[\text{Cov}_{\ell}]^{-1}_{XY} \frac{\partial C_{X \ell}}{\partial p_j},
\ee
where $C_{X \ell}$ is the power in the $\ell^{th}$ multipole for $X= T,E, B$ given by
\be
[\text{Cov}_{\ell}]_{XX} = \frac{2}{(2\ell+1) f_{sky}} \bra{C_{X \ell} +N_{\ell}}
\ee where $N_{\ell}$, the noise level, depends on the data type. The noise is specified by the experiment. 

Because there is a strict geometric degeneracy between $\Omega_\Lambda$, $w_0$ and $w_a$, 
finding the derivatives of the dark energy equation of state (EOS) parameters while keeping $\Omega_\Lambda$ 
fixed artificially breaks this degeneracy. To this end, we follow \cite{DETF} and 
start with computing the Fisher matrix for the CMB with the following parameters: 
p=$\brac{\omega_b, \omega_c, \theta_s, \tau, n_s, A_s}$ 
where $\theta_s$ is the angular size of the sound horizon. This can be written as
\be
\theta_s = \pi \frac{r_s(z_{cmb})}{r(z_{cmb})},
\ee where $r_s(z_{cmb})$ is the sound horizon given in equation \ref{soundhorizon} and 
$r(z_{cmb})$ is the comoving distance to the last scattering surface
\be
r(z) = c\int^{z}_{0} \frac{1}{H(z')} dz'.
\label{sound horizon}
\ee
To compute the derivative $\partial C_{\ell}/\partial \theta_s$, we use the transformation
\be
\frac{\partial C_{\ell}}{\partial \theta_s } \simeq \frac{\Delta C_{\ell}}{\Delta \Omega_{\Lambda}} \frac{\Delta \Omega_{\Lambda}}{\Delta \theta_s },
\ee 
and when evaluating $\frac{\Delta \theta_s}{\Delta \Omega_\Lambda }$, $\omega_c$ and $\omega_b$ must stay fixed by compensating with $h$ through
\be
h^2 = \frac{\omega_b+\omega_c}{1-\Omega_\Lambda}.
\ee

The resulting Fisher matrix $F$ is then transformed back into $\tilde{F}$, corresponding to the parameters p'=$\brac{\omega_b, \omega_c, \Omega_\Lambda, \tau, n_s, A_s, w_0, w_a}$ using
\be
\tilde{F}_{ij} = \sum_{n,m} \frac{\partial p_m }{\partial p_i} F_{mn} \frac{\partial p_n }{\partial p_j}.
\ee
The non-trivial expressions needed for the Jacobian are derivatives of $\theta_s$ with respect to $w_0$, $w_a$, $\Omega_\Lambda$, $\omega_c$ and $\omega_b$.\\
 
For the calculations of the derivatives of the power spectrum with respect to the isocurvature amplitudes  in equation \ref{eq:FM}, we have adopted the treatment in \cite{Bucher02} where the pure isocurvature modes are normalized to have the same power in their CMB temperature spectra as the adiabatic model. This normalization is applied to both the CMB and LSS spectra. 

\subsection{The impact of isocurvature modes on dark energy}
In this section we consider the impact of admitting isocurvature initial conditions on the constraints on the dark energy parameters.
We follow \cite{DETF} and choose not to focus on the constraints on $\Omega_X$.

\begin{table}
\begin{center}
\begin{tabular}{llllll}
\hline
\hline
Experiment & {\planck}& & &  {\planck} & \\ 
 & + {\BOSS} & & & +{\ADEPT} & \\ 
\hline
Parameters & $\sigma(w_0)$ & $\sigma(w_a)$ && $\sigma(w_0)$ & $\sigma(w_a)$ \\
\hline
Adiabatic mode    & 0.045  & 0.083 && 0.035 & 0.060 \\
\hline
Adiabatic + 1 ISO mode  & && & & \\
\hline
{\adm}+{\cim}+{\adci} & 0.058  & 0.087 && 0.042  & 0.063  \\
{\adm}+{\nidm}+{\adnid}& 0.048 & 0.086 && 0.038  &  0.062 \\ 
{\adm}+{\nivm}+{\adniv} & 0.049  & 0.091  && 0.037  & 0.070 \\
\hline
Adiabatic + 2 ISO modes & && & & \\
\hline
{\adm}+{\cim}+{\nidm}+corr &0.058 &0.089  && 0.043 & 0.065 \\
{\adm}+{\cim}+{\nivm}+corr  & 0.069 & 0.11  && 0.055 & 0.090 \\ 
{\adm}+{\nivm}+{\nidm}+corr & 0.055 & 0.098 && 0.043 & 0.080 \\
\hline
Adiabatic + all ISO modes & 0.073& 0.12  && 0.061 & 0.10 \\
\hline 
\hline 
\end{tabular}
\end{center}
\caption{Table summarizing the constraints on $(w_0,w_a)$ for adiabatic and admixtures of uncorrelated adiabatic and isocurvature modes, marginalizing other all other parameters, for the {\BOSS} and {\ADEPT} experiments. The decrease in the figure of merit when all isocurvature modes are added to the adiabatic case are FoM$_{ISO}$=0.54 FoM$_{AD}$ and FoM$_{ISO}=0.47$ FoM$_{AD}$ for {\BOSS}$+${\planck} and {\ADEPT}$+${\planck} respectively. The fiducial model assumes adiabaticity.}
\label{table2}
\end{table}

We compute the potential errors on $w_0$ and $w_a$ for different subsets of adiabatic and isocurvature 
initial conditions while marginalizing over all other cosmological parameters. The results for both 
the {\BOSS} and {\ADEPT} experiments are summarized in table \ref{table2}. We find a systematic degradation
of the viable constraints on dark energy as more degrees of freedom are added. 
In order to quantify the constraining power of the data, we compute the Dark Energy Task Force (DETF) Figure of merit (FoM), which is defined as the reciprocal of the area in
the $w_0 - w_a$ plane, enclosing the $95\%$ confidence limit (CL) region \cite{DETF}. 
We are concerned with the change in the FoM when isocurvature modes are introduced 
relative to the case of pure adiabaticity. The {\BOSS} FoM is found to decrease by $46\%$ from pure adiabaticity to the case in which all isocurvature 
modes are admitted in addition to the adiabatic, while the {\ADEPT} FoM degrades by $53\%$. We note that the results quoted here are slightly different to those quoted in \cite{ibao1}. This is due to the different normalization method used in \cite{ibao1}, which follows \cite{Moodley04}, whereas we follow \cite{Bucher02}. However, the different normalization methods used have little impact on the results reported here, which display a similar trend to \cite{ibao1} but with a slightly larger relative degradation of parameter errors when isocurvature modes are included.

\begin{figure}[hb!]
\centering
\subfigure[$\mbox{ C}_{\ell}^{TT}$ spectrum]{
\label{cmb_breaking1}
{\includegraphics[scale=0.6]{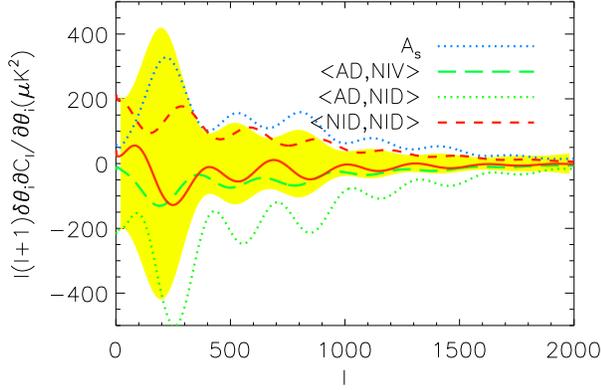}}}
\subfigure[$\mbox{ P(k)}$ at $z=0.35$]{
\label{cmb_breaking}
{\includegraphics[scale=0.6]{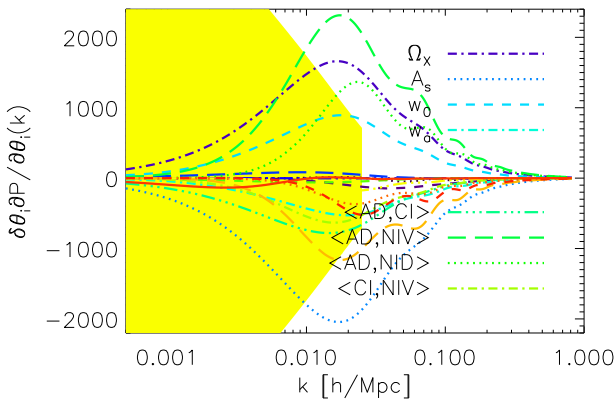}}}
\caption{Main contributions to the degenerate direction with the highest isocurvature fraction \textbf{(a)} in the CMB data from {\planck} alone, \textbf{(b)} in the matter power spectrum using the {\BOSS} dataset alone. The red solid line is the total derivative in the considered degenerate direction, which cancels to within the limits allowed by {\planck} and {\BOSS} error bars (yellow region). For the matter power spectrum, we only plot the yellow region up to $k_{max}=0.1\mbox{ h Mpc}^{-1}$ in this redshift bin.}
\label{degendirection}
\end{figure}

\begin{figure}[ht!]
\centering
\subfigure[$\mbox{ }\xi(r)$ degeneracy]{
\label{xi_boss_alone}
{\includegraphics[scale=0.6]{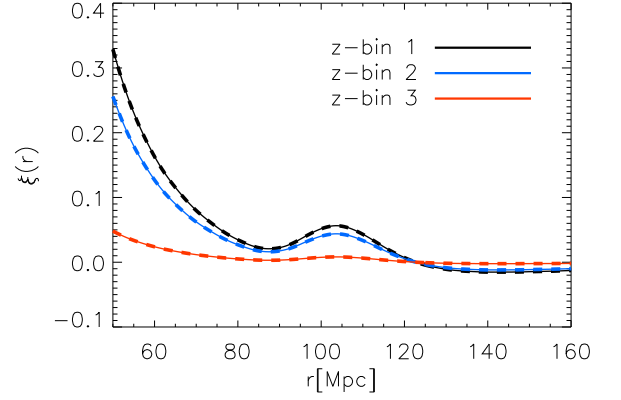}}}
\subfigure[$\mbox{ }\delta \xi /\xi$]{
\label{xi_boss_alone2}
{\includegraphics[scale=0.6]{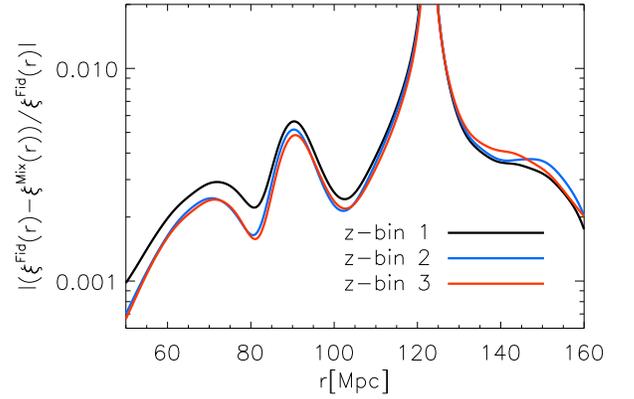}}}
\caption{\textbf{(a)} The galaxy correlation functions for the {\BOSS} survey for different redshift bins. Solid lines represent the purely adiabatic $\Lambda$CDM fiducial model while dashed lines represent a mixed model with $f_{\ISO}=95\%$, $w_0=-1.07$ \& $w_a=-0.133$.  $z=0.35$ (black), $z=0.6$ (blue) and $z=3$ (red). \textbf{(b)} Relative difference of the fiducial and mixed models.}
\label{xi_boss_alone_all}
\end{figure}

The results suggest that no single mode in particular and its correlation are responsible for the change in the allowable $(w_0,w_a)$ region, but rather a mixture of all extra 
degrees of freedom. In order to determine the combination of parameters that is responsible for this degradation, we diagonalize the full 17x17 Fisher matrix corresponding to the {\planck} and LSS datasets separately and find the eigenvector with the smallest eigenvalue, corresponding to the direction that is least constrained by the data, that is, we consider the worst case degenerate scenario. Considering the {\planck} data alone and discarding the degeneracy in the $w_0-w_a$ direction which is the main degenerate direction, four parameters, namely $A_s$, $A_{\adnivE}$, $A_{\adnidE}$ and $A_{\nidE}$ define the most degenerate direction involving isocurvature modes, with the impact of the scalar amplitude being compensated for by a combination of the cross-correlated modes and the optical depth. The degenerate direction in the LSS data (using {\BOSS} as an example) is more complicated and involves a combination of isocurvature parameters and dark energy parameters, namely $\Omega_X$, $A_s$, $w_0$, $w_a$, $A_{\adciE}$, $A_{\adnivE}$, $A_{\adnidE}$ and $A_{\cinivE}$. Figure \ref{degendirection} shows how the perturbations in the different parameters contribute to the the total change in the CMB power spectrum and the matter power spectrum.

The total derivative (shown in red) lies within the noise limits of the respective experiments, making the net change undetectable by the data.\\

\begin{table*}[ht!]
\begin{center}
\begin{tabular}{llllll}
\hline
\hline
Experiment & {\planck} + {\BOSS} & & & {\planck} + {\ADEPT} & \\ 
\hline
Parameters &  $\delta w_0$ & $\delta w_a$ && $\delta w_0 $ & $\delta w_a $ \\
\hline
Adiabatic + 1 ISO mode & && & & \\
\hline
{\adm}+{\cim}+{\adci} & -0.12 (2.7) & 0.077 (0.9) && -0.076 (2.2) & 0.072 (1.2) \\
{\adm}+{\nidm}+{\adnid}& -0.060 (1.3)  & 0.084 (1.0) && 0.050 (1.4) & -0.067 (1.1) \\ 
{\adm}+{\nivm}+{\adniv} & 0.091 (2.0)  & -0.17 (2.0) && -0.055 (1.6)  &  0.16 (2.7) \\
\hline
Adiabatic + 2 ISO modes & && & & \\
\hline
{\adm}+{\cim}+{\nidm}+corr & -0.11 (2.4) & 0.094 (1.1) && -0.075 (2.1) &  0.073 (1.2) \\
{\adm}+{\cim}+{\nivm}+corr  & -0.19 (4.2) & 0.22 (2.7) && 0.16 (4.6)  & -0.23 (3.8) \\ 
{\adm}+{\nivm}+{\nidm}+corr & -0.11 (2.4) & 0.19 (2.3) && -0.088 (2.5) & 0.20 (3.3)  \\
\hline
Adiabatic + all ISO modes & 0.22 (4.9) & -0.29 (3.5)  &&  0.20 (5.7)  & -0.29 (4.8)  \\
\hline 
\hline 
\end{tabular}
\end{center}
\caption{Table summarizing the biases on $(w_0,w_a)$ that could arise from the incorrect assumption of adiabatic initial
conditions, given a universe with an admixtures of uncorrelated 
adiabatic and isocurvature modes for the {\BOSS} and {\ADEPT} experiments. The quantities in brackets are the biases, quoted in number of $1\sigma$ error bars corresponding to the case when
pure adiabaticity is assumed.}
\label{biases}
\end{table*}


Clearly the dark energy model is degenerate with the particular combination of isocurvature modes in the BAO data. The implication is that the constraints on dark energy are at risk of being substantially biased if adiabaticity is incorrectly assumed. 
To emphasize this point, Figure \ref{xi_boss_alone} compares the correlation function, defined by

\begin{equation}
\xi(r)=\int_{0}^{\infty} k^2 P(k) \frac{\sin{kr}}{kr}dk,
\label{eq2}
\end{equation}
where $P(k)$ is the matter power spectrum, that would be measured today for our fiducial $\Lambda$CDM model assuming pure adiabaticity, to a cosmological model assuming dynamical dark energy, described by $w_0=-1.07$ and $w_a=-0.133$, and an admixture of initial conditions, $95\%$ of which is isocurvature in nature. The correlation function is degenerate in all three redshift bins of the {\BOSS} experiment. Note that this degeneracy is completely broken by the CMB data. We now wish to quantify this bias.

For a Gaussian-distributed likelihood function, it can be shown that the linear bias in a set of parameters that we wish to constrain, $\delta \theta_i,$ due to erroneous values of a set of fixed parameters, $\delta \phi_j,$ is  \cite{Taylor}
\be
\delta \theta_i = -\bras{F^{\theta \theta}}_{im}^{-1} F^{\theta \phi}_{mj}\delta \phi_j
\ee where $F^{\theta \theta}$ is the Fisher sub-matrix for the parameters we wish to constrain and
$F^{\theta \phi}$ is a Fisher sub-matrix constructed from the product of the derivatives of the
power spectrum with respect to the parameters being constrained and those which are being fixed. In our case $j$ labels the isocurvature mode amplitudes, incorrectly fixed to zero, $m$ 
labels the eight cosmological parameters that are biased, and $i$ labels the subset of two 
dark energy parameters whose bias is of interest to us.
In order to set $\delta \phi_j$, we diagonalize the combined {\planck} and large-scale 
structure (LSS) Fisher matrix and select the eigenvector, ${\bf e}_i$ with the
smallest eigenvalue $\lambda_i$. This corresponds to the direction in parameter space 
which is least constrained by the data.  
We then take $\delta \phi_j = \sqrt{\frac{M}{\lambda_j}}{\bf e}_j$, where M depends on the total number of cosmological and isocurvature parameters.

We first consider the case of an admixture of the adiabatic mode and the CDM isocurvature mode. For this case we find the biases in the dark energy parameters to be $\delta w_0= -0.12$ and  $\delta w_a = 0.077$ for the {\BOSS} experiment. Comparing the mean biases to the $1\sigma$ 
constraints obtained when pure adiabaticity is assumed, we find that neglecting this 
isocurvature contribution leads to a $2.7\sigma$ and $0.9\sigma$ error in the dark energy parameter estimates for $w_0$ and $w_a$ respectively, when compared to the error forecasts assuming adiabaticity.
If we repeat the calculation for the more advanced experiment {\ADEPT}, we find $\delta w_0= -0.076$ and  
$\delta w_a = 0.072$, equivalent to $2.2\sigma$ and $1.2\sigma$ errors in the dark energy parameters respectively when compared to the adiabatic constraints.

Although no theoretical models for generating the neutrino isocurvature models have thus far been proposed, we would like to conduct a comprehensive exploration of the impact of the initial conditions on the BAO constraints and therefore admit all possible isocurvature degrees of freedom. Table \ref{biases} summarizes the biases for different admixtures of adiabatic and subsets of isocurvature modes. The results are consistent with the degenerate directions in parameter space identified earlier.  For example,  the admittance of the CI and NIV isocurvature modes and their cross correlations has the potential to cause a bias in the dark energy parameters by as much as $4.2\sigma$, in the case of {\BOSS}.

For the case of an admixture of adiabatic and all isocurvature modes and their cross correlations, 
we find that the biases are $\delta w_0= 0.22$ and  $\delta w_a = -0.29$ for the {\BOSS} experiment.  This means that if the initial conditions of our universe are comprised of a 
sub-dominant contribution from all isocurvature modes (within the $1\sigma$ constraints 
from the {\planck} and {\BOSS} experiments), 
the assumption of adiabaticity could lead to an incorrect $4.9\sigma$ detection of non-$\Lambda$ 
dark energy model or a $3.5\sigma$ false claim of dynamics.  Alternatively, 
$\Lambda$ could be found to be consistent with the data when in 
fact $w(z)\neq -1$.  The potential bias incurred by the adiabatic assumption 
in the case of the {\ADEPT} experiment has a mean of $\delta w_0= 0.20$ (equivalent to $5.7\sigma$) while the measurement of $w_a$ could be inaccurate at the level of only $4.8\sigma$.

\subsubsection{Constraints on isocurvature modes from the LSS data}
We now consider the impact of the large scale structure information on isocurvature constraints. Although allowing for isocurvature modes degrades the dark 
energy constraints relative to the pure adiabatic case, 
this analysis has revealed a powerful positive. The forecasted errors on the isocurvature parameters based on 
the CMB data alone and in conjunction with the LSS experiments are compared in tables 
\ref{forecasts:tab:constraints:single}, \ref{forecasts:tab:constraints:double} and \ref{forecasts:tab:constraints:full} respectively for single, double and fully correlated isocurvature modes. We find that the error bars on the isocurvature parameters decrease by a few percent to as much as $96\%$ for certain modes when the LSS data (either {\BOSS} or {\ADEPT}) is added to the {\planck} data. Assuming an adiabatic fiducial model, the measurement of the BAO in the first redshift bin of the {\BOSS} experiment and the CMB by {\planck} will reduce the allowed isocurvature fraction from $5.6\%$ for the CMB data only to $3.5\%$, and to $3.1\%$ and $3\%$ when adding the information from LSS in the second and the third redshift bins. 

\begin{table}[ht!]
\begin{center}
\begin{tabular}{lll}
\hline
\hline
& Adiabatic + 1 ISO mode &\\
\hline
 & {\planck} + {\BOSS} &  {\planck} + {\ADEPT}\\ 
\hline
{\ad} & 0.31 (4) & 0.32 (1) \\
{\ci} & 0.012 (5) & 0.012 (6) \\
{\adci} & 0.035 (96) & 0.032 (96) \\
\hline
{\ad} & 0.027 (14) & 0.26 (18)  \\
{\nid} & 0.0050 (6) & 0.0049 (9) \\
{\adnid} & 0.016 (13) & 0.015 (18) \\
\hline
{\ad} & 0.25 (21) & 0.24 (24) \\
{\niv} &  0.0089 (5) & 0.0089 (6) \\
{\adniv} & 0.090 (90) & 0.057 (94) \\
\hline 
\hline
\end{tabular}
\end{center}
\caption{Forecasted uncertainties on isocurvature parameters for different cases for the {\planck} and LSS data ({\BOSS} and {\ADEPT}) for single isocurvature modes. The percentage improvement in $1\sigma$ errors when the LSS data is added to the {\planck} data is shown in brackets.}
\label{forecasts:tab:constraints:single}
\end{table}

\begin{table}[h!]
\begin{center}
\begin{tabular}{lll}
\hline
\hline
& Adiabatic + 2 ISO modes &\\
\hline
 & {\planck} + {\BOSS} &  {\planck} + {\ADEPT}\\ 
\hline
{\ad} & 0.36 (19) & 0.34 (24)\\
{\ci} & 0.20 (5) & 0.020 (7) \\
{\nid} & 0.0072 (2) &  0.0071 (3) \\
{\adci} & 0.035 (96) & 0.032 (96) \\
{\adnid} & 0.029 (8) & 0.027 (13) \\
{\cinid} & 0.015 (2) & 0.015 (3)  \\
\hline
{\ad} & 0.43 (4) & 0.43 (4) \\
{\ci} & 0.017 (5) & 0.017 (5)  \\
{\niv} & 0.0095 (4) & 0.0095 (4) \\
{\adci} & 0.044 (96) & 0.043 (96) \\
{\adniv} & 0.096 (91) & 0.068 (94) \\
{\ciniv} & 0.017 (3) & 0.017 (3) \\
\hline
{\ad} & 0.28 (38)  & 0.27 (40)  \\
{\nid} & 0.0083 (3) & 0.0082 (4) \\
{\niv} & 0.018 (3) & 0.018 (3) \\
{\adnid} & 0.028 (10) & 0.028 (11) \\
{\adniv} & 0.093 (90) & 0.062 (93) \\
{\nidniv} & 0.016 (3) & 0.016 (4)  \\
\hline 
\hline
\end{tabular}
\end{center}
\caption{Forecasted uncertainties on isocurvature parameters for different cases for the {\planck} and LSS data ({\BOSS} and {\ADEPT}) for double isocurvature modes. The percentage improvement in $1\sigma$ errors when the LSS data is added to the {\planck} data is shown in brackets.}
\label{forecasts:tab:constraints:double}
\end{table}

\begin{table}[h!]
\begin{center}
\begin{tabular}{lll}
\hline
\hline
& Adiabatic + all ISO modes &\\
\hline
& {\planck} + {\BOSS} &  {\planck} + {\ADEPT}\\ 
\hline
{\ad} &   0.51 (13) & 0.51 (14)  \\
{\ci} &  0.047 (18) & 0.044 (23)  \\
{\nid} &  0.017 (11) & 0.016(14) \\
{\niv} &  0.039 (17)  & 0.037 (22) \\
{\adci} & 0.061(94) & 0.057 (95)   \\
{\adnid} & 0.088 (17) & 0.077 (28)  \\
{\adniv} & 0.10 (91)  &  0.073 (94) \\
{\cinid} &  0.058 (12) & 0.057 (13) \\
{\ciniv} & 0.035 (12) & 0.033 (18) \\
{\nidniv} & 0.042 (6) & 0.042 (6) \\
\hline 
\hline
\end{tabular}
\end{center}
\caption{Forecasted uncertainties on isocurvature parameters for different cases for the {\planck} and LSS data ({\BOSS} and {\ADEPT}) for fully correlated isocurvature case. The percentage improvement in $1\sigma$ errors when the LSS data is added to the {\planck} data is shown in brackets.}
\label{forecasts:tab:constraints:full}
\end{table}

\begin{figure}[h!]
\centering
\subfigure[]{
\label{direction1}
{\includegraphics[scale=0.6]{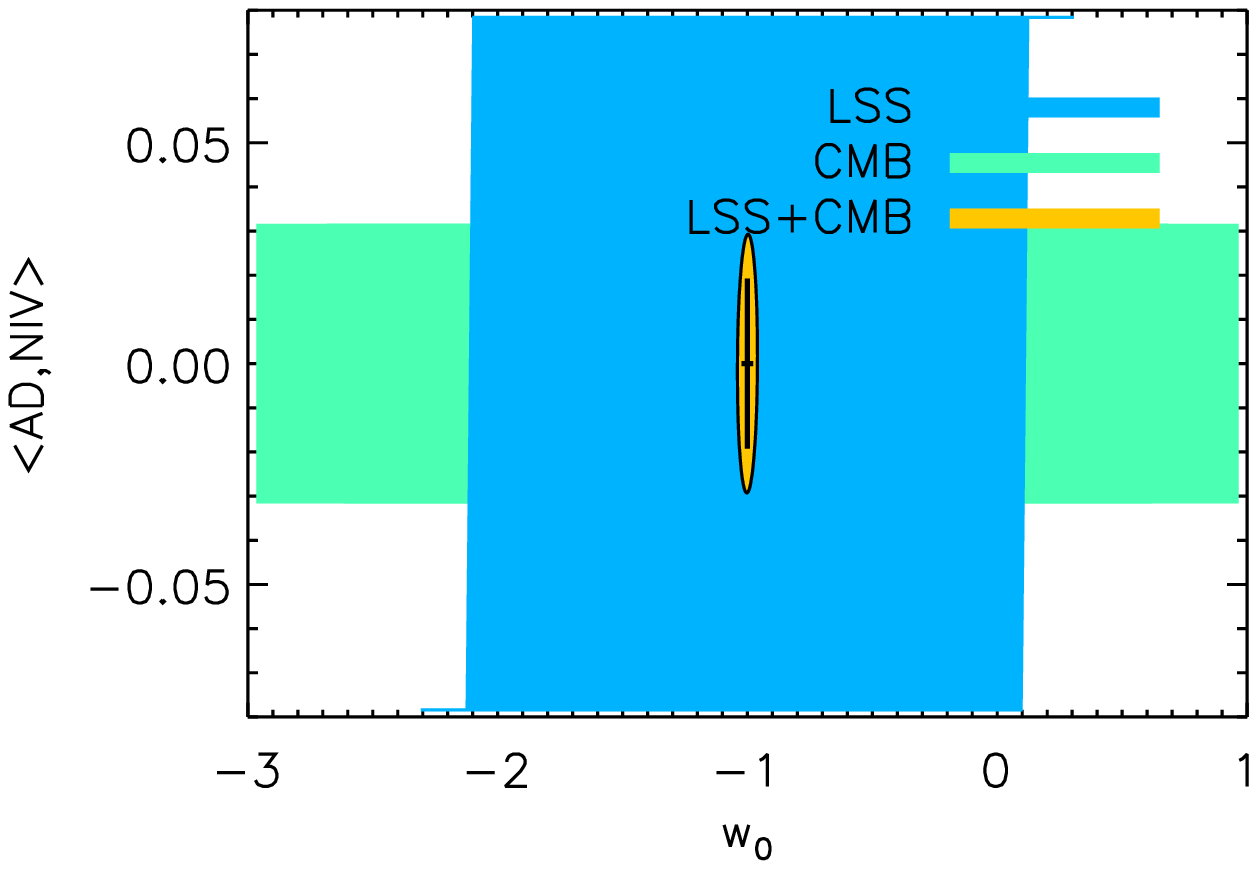}}}
\subfigure[]{
\label{direction2}
{\includegraphics[scale=0.6]{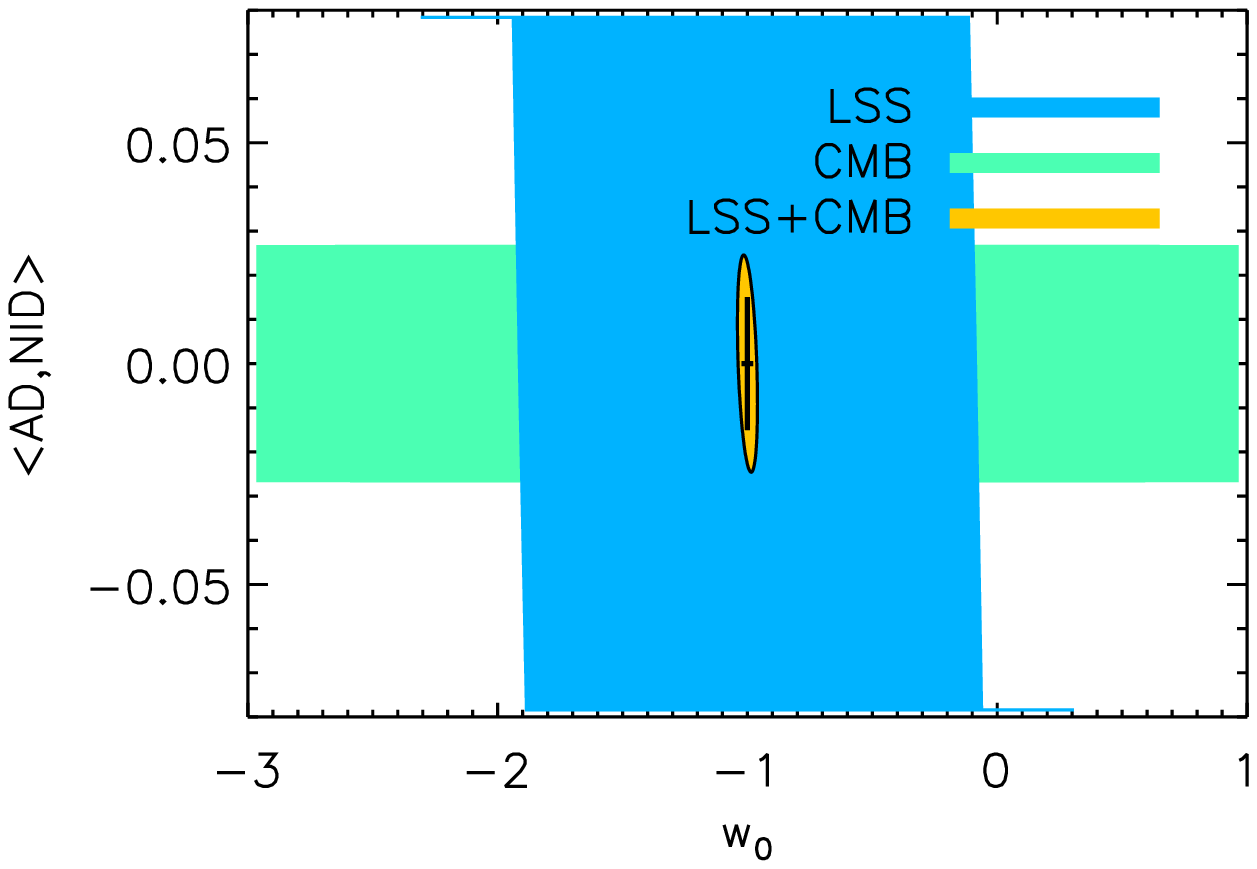}}}
\caption{Effect of combining the CMB and LSS datasets on the $1\sigma$ error ellipses for isocurvature contributions and $w_0$. We have only represented the isocurvature modes ({\adniv} and {\adnid}) that determine the main degenerate direction with the largest isocurvature fraction. Inner straight lines represent the error bars obtained using both CMB and LSS experiments, but assuming a cosmological constant.}
\label{direction}
\end{figure}

The reason for this stems from the fact that the considered degenerate direction in parameter space for the CMB data ({\adniv}, {\adnid}, {\nid}, $A_s$) differs from the degenerate direction of the LSS data ({\adniv}, {\adnid}, $\Omega_X$, $w_0$, $w_a$, $A_s$).

\begin{figure}[ht!]
\centering
\subfigure[]{
\label{adding_bins_1}
{\includegraphics[scale=0.6]{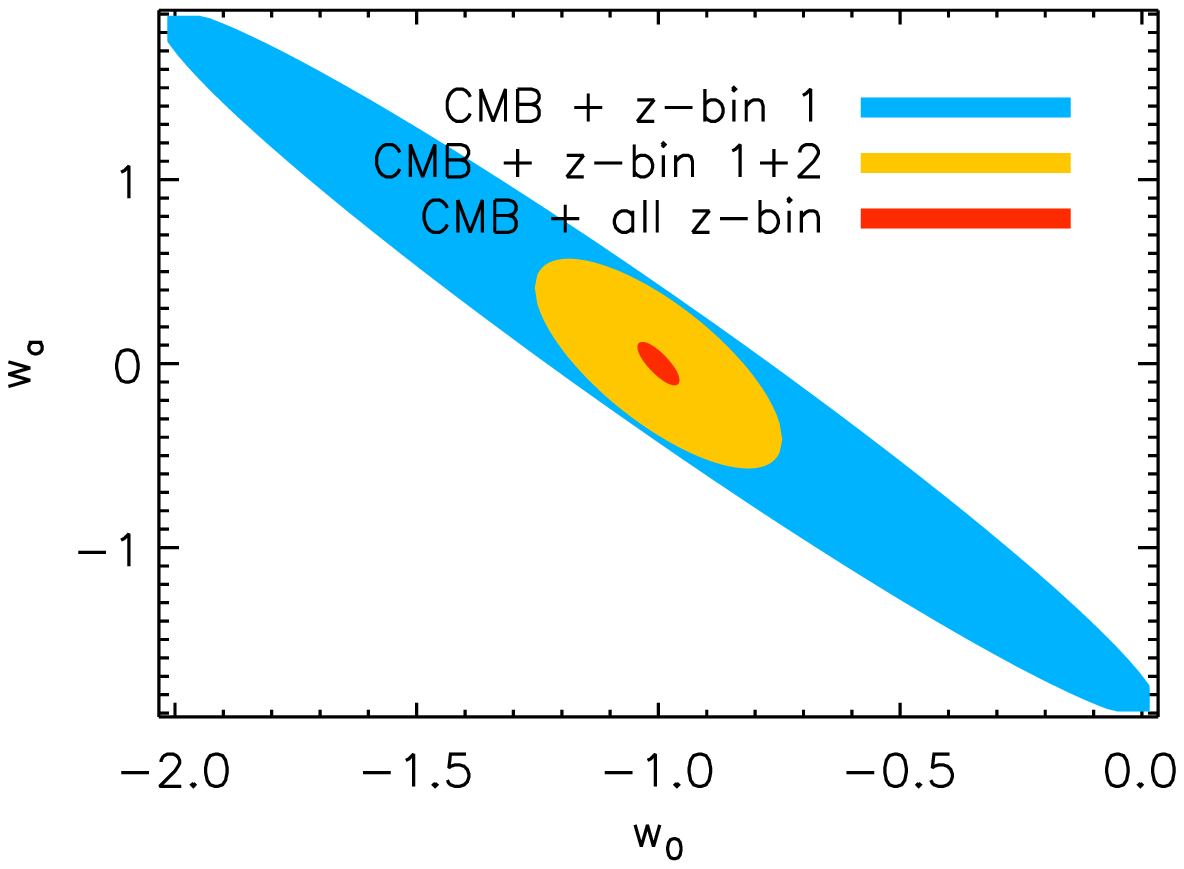}}}
\subfigure[]{
\label{adding_bins_2}
{\includegraphics[scale=0.6]{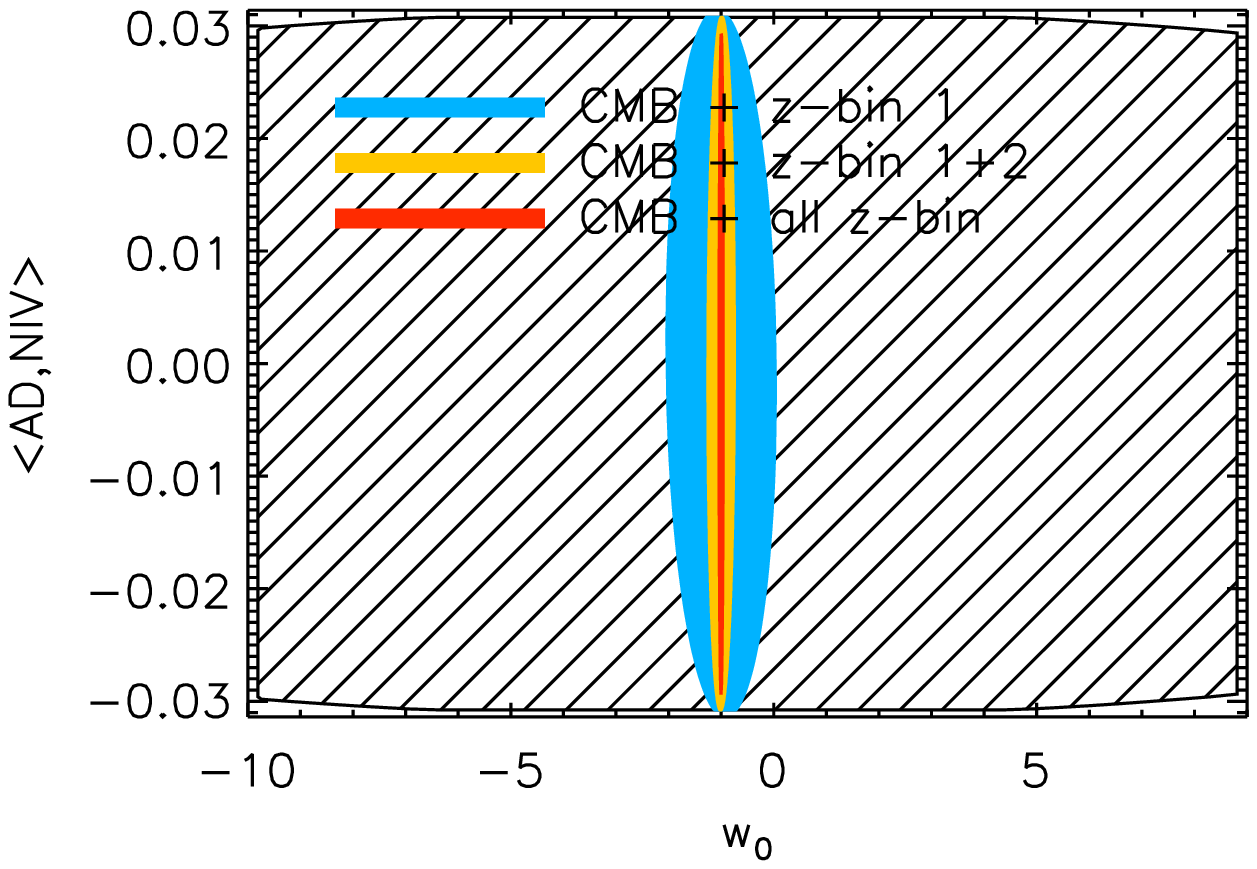}}}
\subfigure[]{
\label{adding_bins_3}
{\includegraphics[scale=0.6]{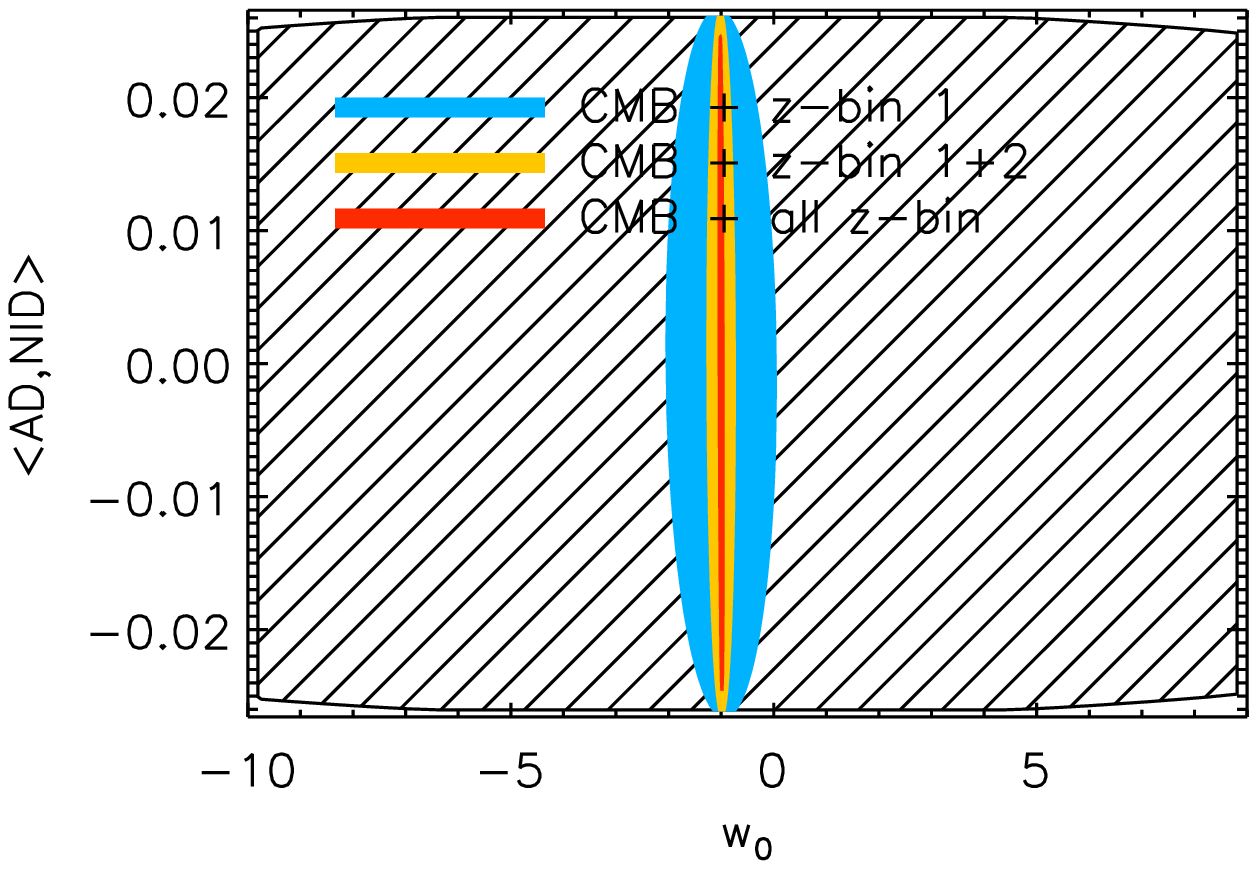}}}
\caption{Effect of adding different redshift bin datasets on the $1\sigma$ error ellipses for $(w_0,w_a)$, $(w_0,{\adnivE})$ $(w_0,{\adnidE})$. Hatched regions on panels (b) and (c) represent the $1\sigma$ error ellipse for the CMB experiment alone.}
\label{adding_bins}
\end{figure}

\begin{figure}[ht!]
\centering
\includegraphics[scale=0.6]{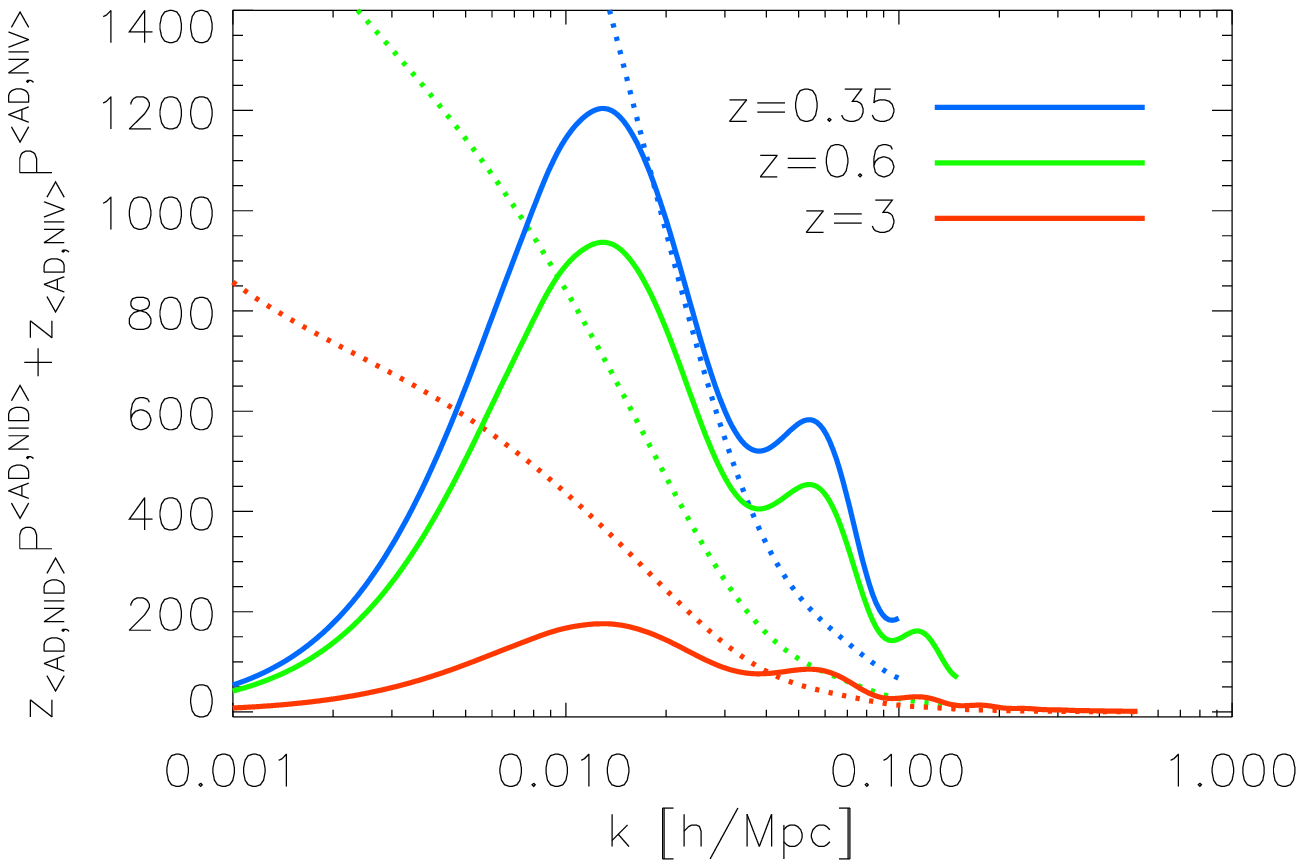}
\caption{Sum of the most dominant isocurvature contributions $z_{\adnivE} P^{\adnivE}+z_{\adnidE} P^{\adnidE}$ to the power spectrum for different redshift bins of the {\BOSS} experiment. The blue, green and red solid curves represents respectively the isocurvature contribution at redshift $z=0.35$, $z=0.6$ and $z=3$. The dotted lines represent the {\BOSS} error bars for different redshift bins.}
\label{corr_ad_nid_ad_niv_z}
\end{figure}

\begin{figure}[ht!]
\begin{center}
\begin{tabular}{c}
\includegraphics[trim = 0mm 0mm 0mm 10mm,scale=0.6, angle=0,width=3.5in,height=1.5in]{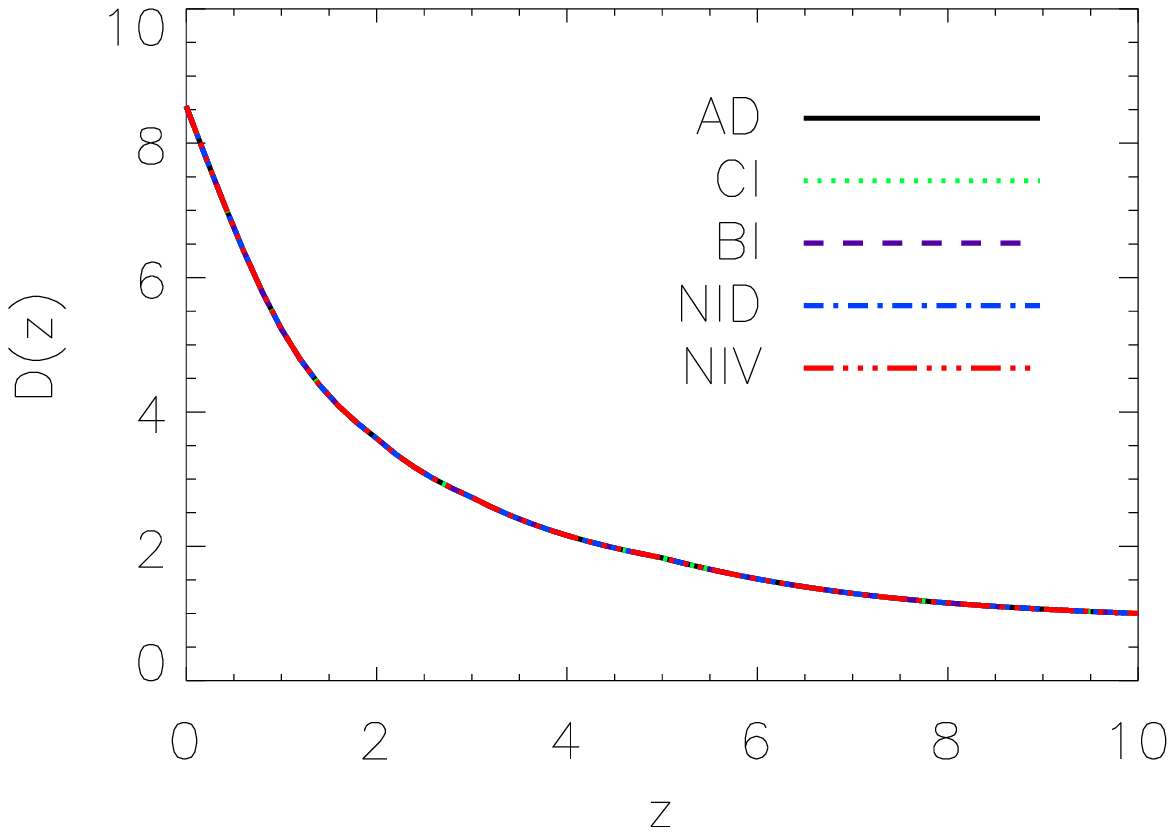}\\ 
\includegraphics[trim = 0mm 0mm 0mm 10mm,scale=0.6, angle=0,width=3.5in,height=1.5in]{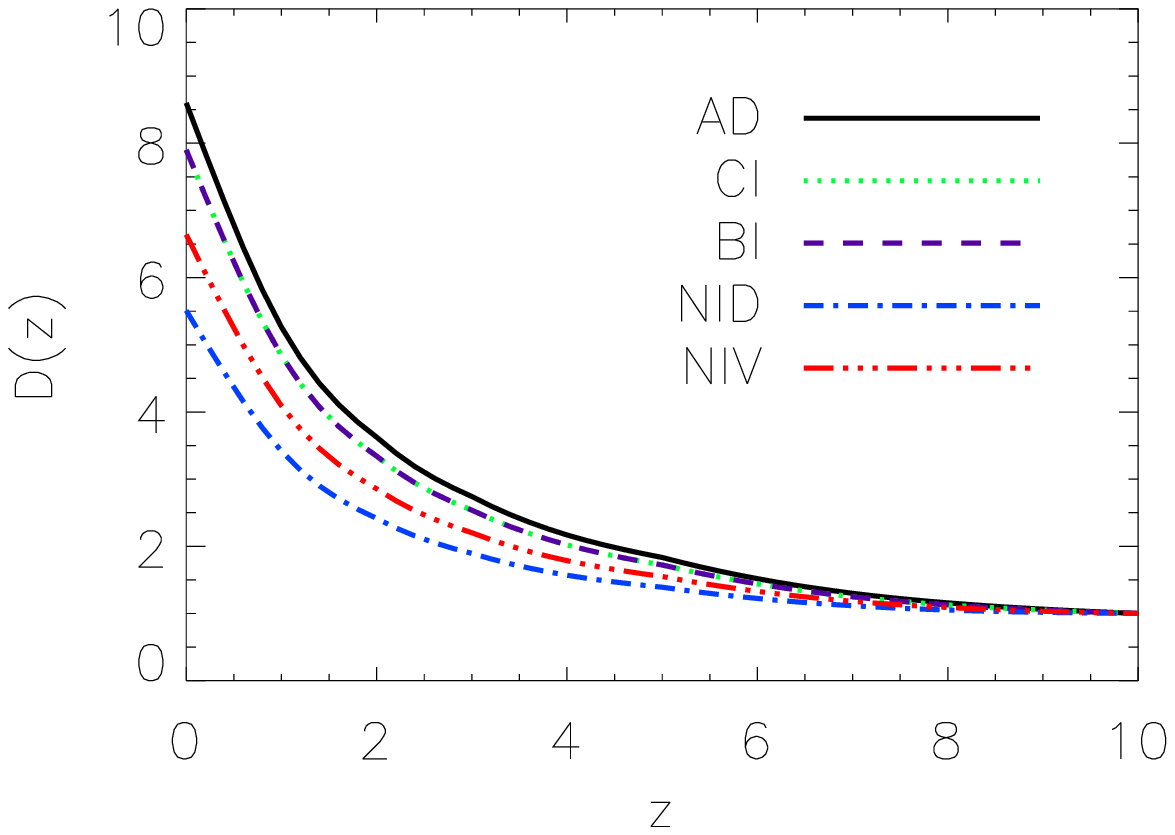}
\end{tabular}
\end{center}
\caption{Growth function as a function of redshift for all modes. \textbf{Top:} on intermediate scales ($k=0.01\mbox{ h Mpc}^{-1}$). \textbf{Bottom:} on large scales $k=5.3\times 10^{-3}\mbox{ h Mpc}^{-1}$. On small and intermediate scales, the growth function is the same for all regular modes, while the AD mode grows faster than the isocurvature modes on large scales. We have normalized the growth to unity at $z=10$.}
\label{growth_func}
\end{figure}

Figure \ref{direction} illustrates the different directions by showing the $1\sigma$ error ellipses for the main isocurvature contributions and $w_0$. The two degenerate directions are almost orthogonal. Here, the inner straight lines represent the marginalised error bars obtained by combining both {\planck} and {\BOSS} experiments in the case of a cosmological constant. Clearly the ability of LSS data to measure isocurvature modes is related to the information provided by the BAO about dark energy. In Figure \ref{adding_bins}, we compare the $1\sigma$ error ellipses for $(w_0,{\adnivE})$ and $(w_0,{\adnidE})$ that are obtained when we include the CMB dataset and add the data from the {\BOSS} redshift bins in succession. We see that the BAO data primarily serves to reduce the phase space for $w_0$ with the largest improvement in the $w_0$ constraint coming from the second redshift bin. As the redshift increases, the contribution from dark energy diminishes until matter comes to dominate, at which time the impact of dark energy on the observables is small. For this reason, the intermediate redshift bin for {\BOSS} centered at $z=0.6$ provides the best constraints on $w_0$.  In figure \ref{corr_ad_nid_ad_niv_z} we compare the sum of the most dominant isocurvature contributions $z_{\adnivE} P^{\adnivE}+z_{\adnidE} P^{\adnidE}$ to the power spectrum at the different redshift bins of the {\BOSS} experiment to their respective error bars.  The area between the solid (signal) and dotted (error) curves indicates the amount of information provided by each bin.  Clearly, this combination of isocurvature parameters is best constrained from the measurement of the galaxy power spectrum at $z=0.6$ for this particular experiment. Furthermore, the differing shapes of the signal curves suggests that complementary information is available at different redshifts.  Hence, the measurement of the BAO scale at different redshifts between decoupling and today helps to constrain the isocurvature modes. 

We note that in this study we have assumed exact knowledge of the nonlinear shift of the BAO location as a function of redshift.

As an aside, we note that information about the initial conditions from LSS data does not stem from differences in the growth rates for different modes. Figure \ref{growth_func} shows the growth function of the perturbations on intermediate (top) and on large scales (bottom). It is clear that on very large scales, the isocurvature modes grow more slowly than the adiabatic modes. This is expected as perturbations which are isocurvature in nature only grow when they enter the horizon while adiabatic fluctuations grow at all times. However on the scales probed by the BAO signal, the isocurvature modes and adiabatic modes grow at the same rate.

\section{Conclusions}
\label{sec:conclusion}
The first detection of the BAO peak in the galaxy correlation function measured by SDSS opened the door to using the clustering of galaxies on scales of $\sim150\mbox{ Mpc}$ as a cosmic yardstick. By comparing the size of the overdensity of baryons at the epoch of recombination predicted from theory and calibrated by the CMB, with its size as it appears in the large-scale structure of galaxies today, we can study 
the expansion history of the universe. However, in order to succeed in making a precise measurement of the signal we will need the huge volumes probed only by the most recent generation of redshift surveys. With such precision we hope to reveal the nature of dark energy and probe its time evolution if it exists.

With claims of constraints on dark energy from BAO experiments to the level of a few percent, it becomes important to check the assumptions made in the post-observational analysis. In this paper, we have revisited the assumption of pure adiabatic initial conditions and considered the impact of allowing  isocurvature-adiabatic admixtures on the BAO peak and the implications for 
dark energy studies. We have shown that a combination of differences in the baryon growth profile that arises due to the presence of isocurvature modes and Silk damping change both the shape and position 
of the BAO peak. Non-adiabatic initial conditions leave the sound speed unchanged but instead alter the development of the acoustic waves in the baryon-photon fluid prior to decoupling which modifies the scale on which the sound waves imprint on the baryon distribution.

The degeneracy between the impact of mixed initial conditions and the effect of a dynamical dark energy model on the BAO peak weakens the potential constraints on the dark energy parameters forecasted for a combined {\planck} and LSS dataset.  
We found that the admission of more general initial conditions which include isocurvature modes and their cross-correlations increases the $95\%$ confidence region in $(w_0,w_a)$ space by 
$50\%$ in the case of the {\BOSS} experiment and thus the assumption of adiabaticity can lead to the under-estimation of the errors on the dark energy parameters. 
Furthermore, if we assume purely adiabatic initial conditions, we run the risk of attributing a shift in the peak away from the prediction of 
$\sim150$ Mpc for a $\Lambda$CDM universe to a non-$\Lambda$ dark energy model. We have shown that this can lead to a bias in the estimates of the dark energy parameters, leading to a several $\sigma$ incorrect confirmation of $\Lambda$ or detection of non-$\Lambda$. 

On a positive note, the change in the BAO peak in isocurvature models indicates that there is useful information in the galaxy correlation function on the nature of the primordial perturbations even when simultaneously measuring dark energy equation of state parameters. We find that the use of the LSS data in addition to the CMB data substantially improves our ability to measure the contributions of different modes to the initial conditions. The matter power spectrum constrains the dark energy parameters and in so doing breaks the degeneracy in the isocurvature-dark energy parameter space. Furthermore, even when assuming $w=-1$, the degenerate parameter combinations in the CMB and LSS are different.  A similar conclusion is reached in \cite{Carbone} despite differences in their analysis compared with ours, such as their choice of non-adiabatic fiducial model, the inclusion of spatial curvature and a different LSS experiment.

\section{Acknowledgments}
The authors would like to thank D. Eisenstein, H.J. Seo, and A. Rassat for helpful comments. CZ was funded by a NRF/DST (SA) Innovation Fellowship and a National Science Foundation (USA) fellowship under grant PIRE-0507768. PO is funded by the SKA (SA). SMK received support via the Meraka Institute via funding for the South African Centre for High Performance Computing (CHPC). BB and KM acknowledge support from the National Research Foundation, SA.

\section{Appendix}

The derivatives of the matter power spectrum with respect to the isocurvature parameters have not been presented in the literature before and we show them here for the benefit of the reader, in addition to the derivatives with respect to the cosmological parameters.

 \begin{figure*}[ht!]
 \begin{center}
 \begin{tabular}{ccc}
  \includegraphics[trim = 0mm 0mm 0mm 10mm, scale=0.37, angle=0]{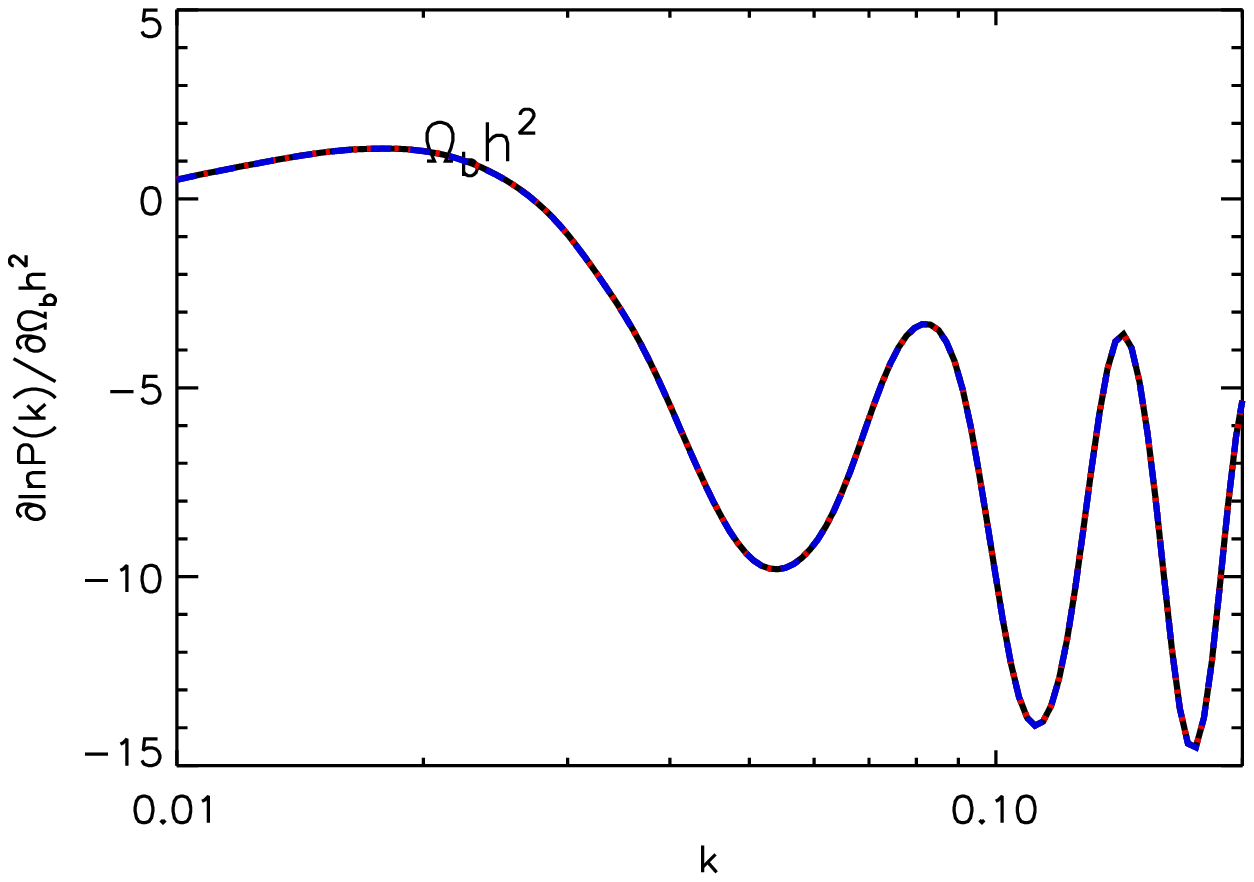} 
 &\includegraphics[trim = 0mm 0mm 0mm 10mm, scale=0.37, angle=0]{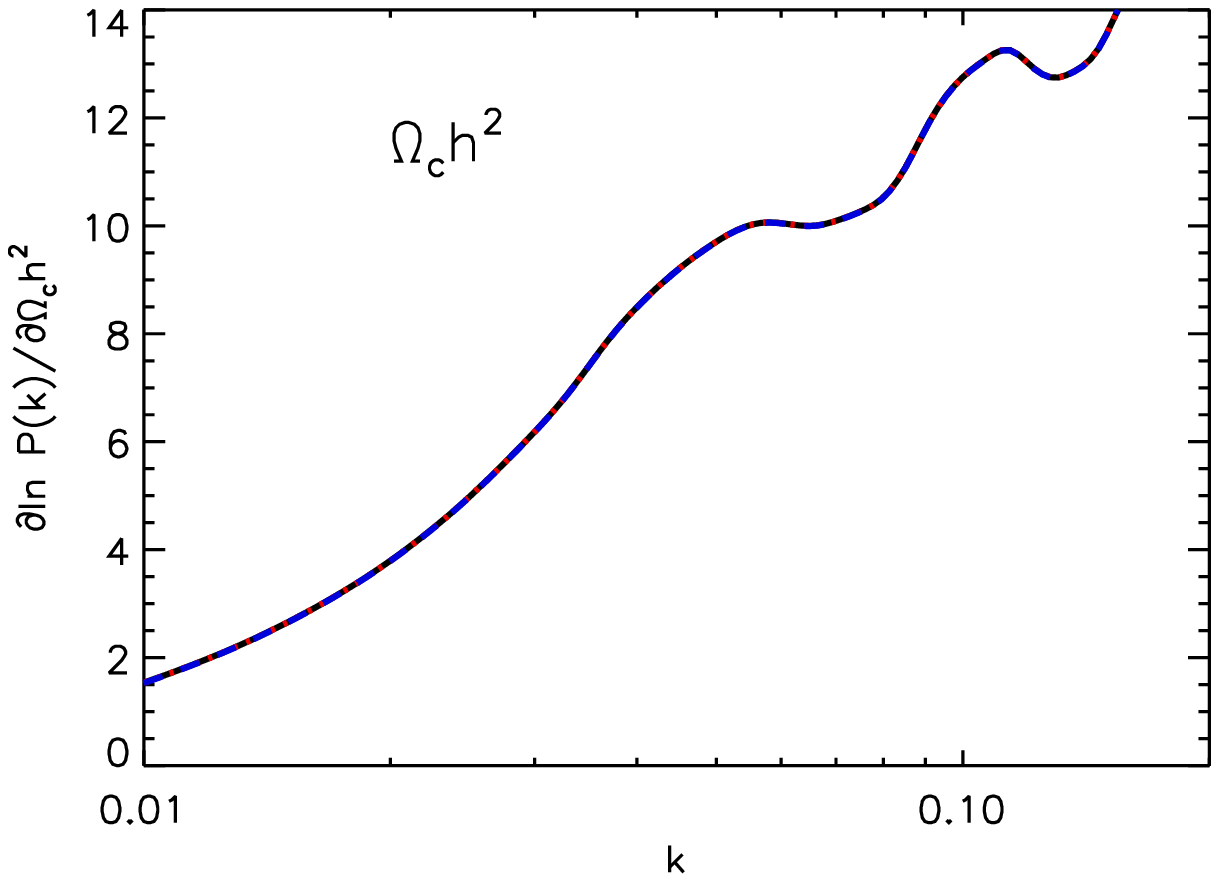}
 &\includegraphics[trim = 0mm 0mm 0mm 10mm, scale=0.37, angle=0]{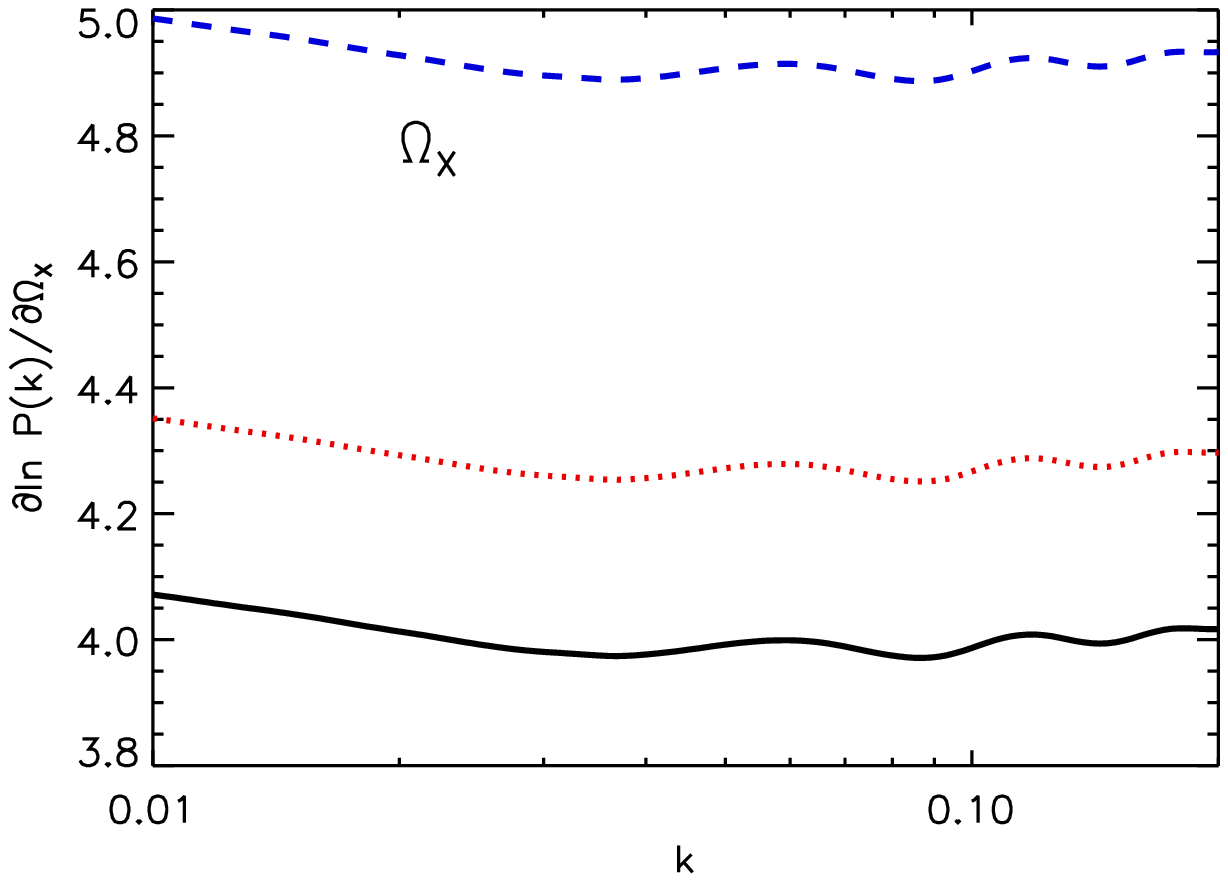}\\ 
  \includegraphics[trim = 0mm 0mm 0mm 10mm, scale=0.37, angle=0]{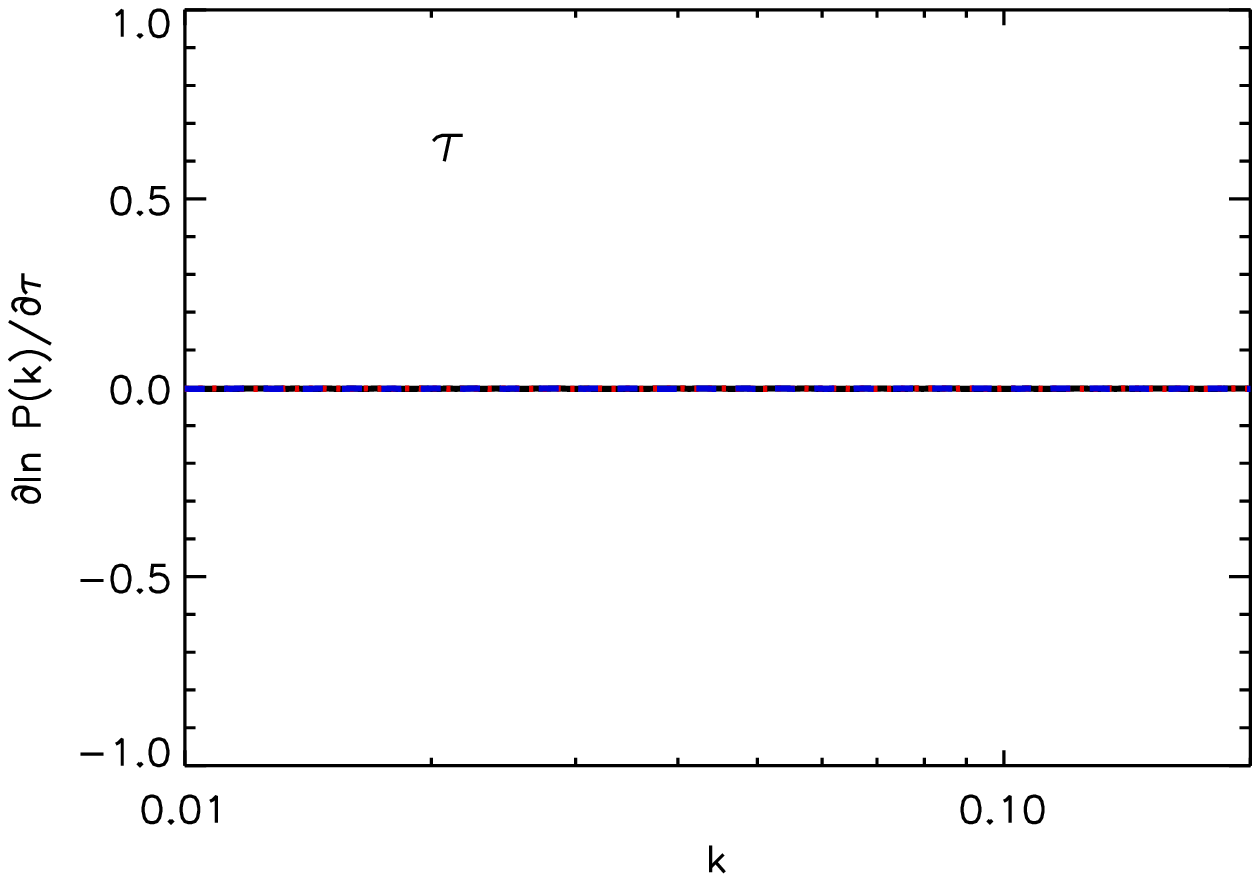}
 &\includegraphics[trim = 0mm 0mm 0mm 10mm, scale=0.37, angle=0]{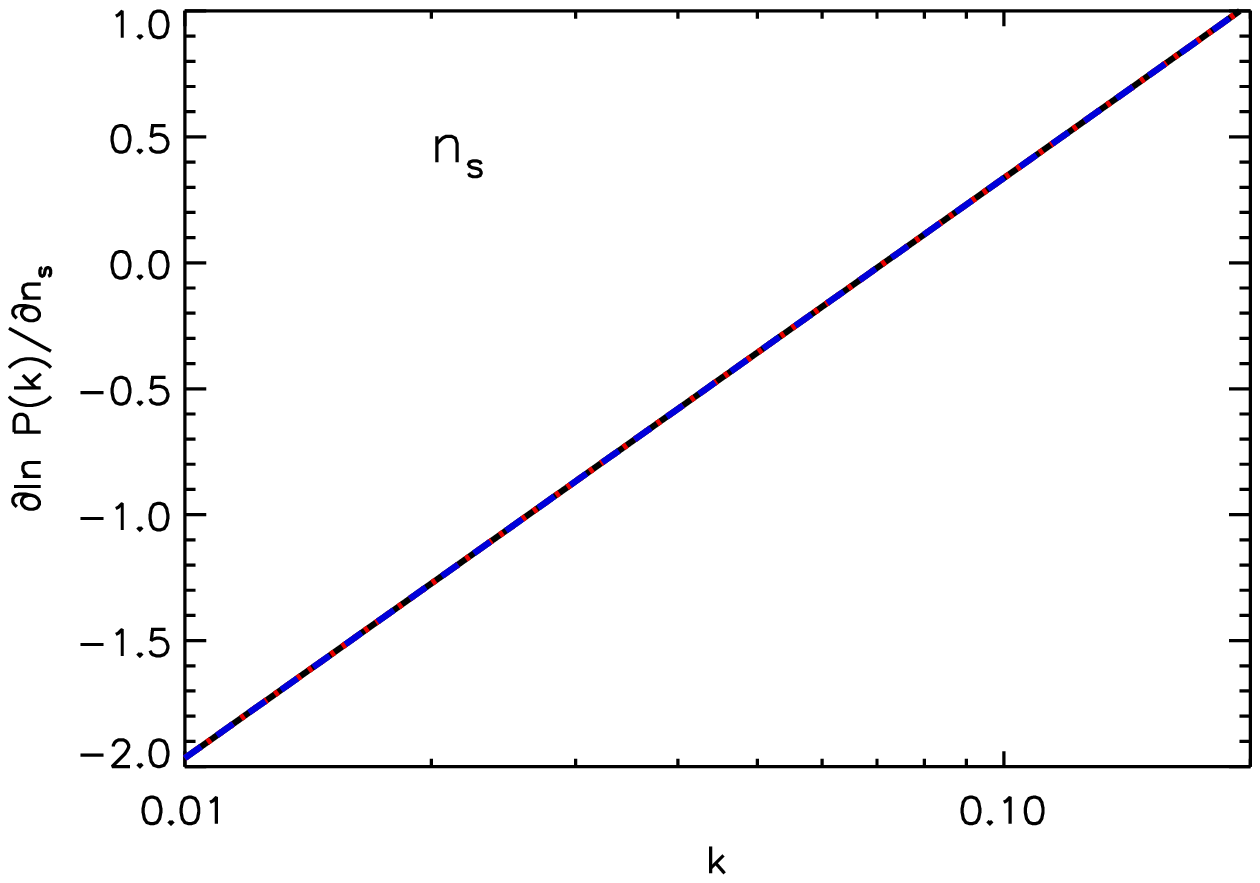}
 & \includegraphics[trim = 0mm 0mm 0mm 10mm, scale=0.37, angle=0]{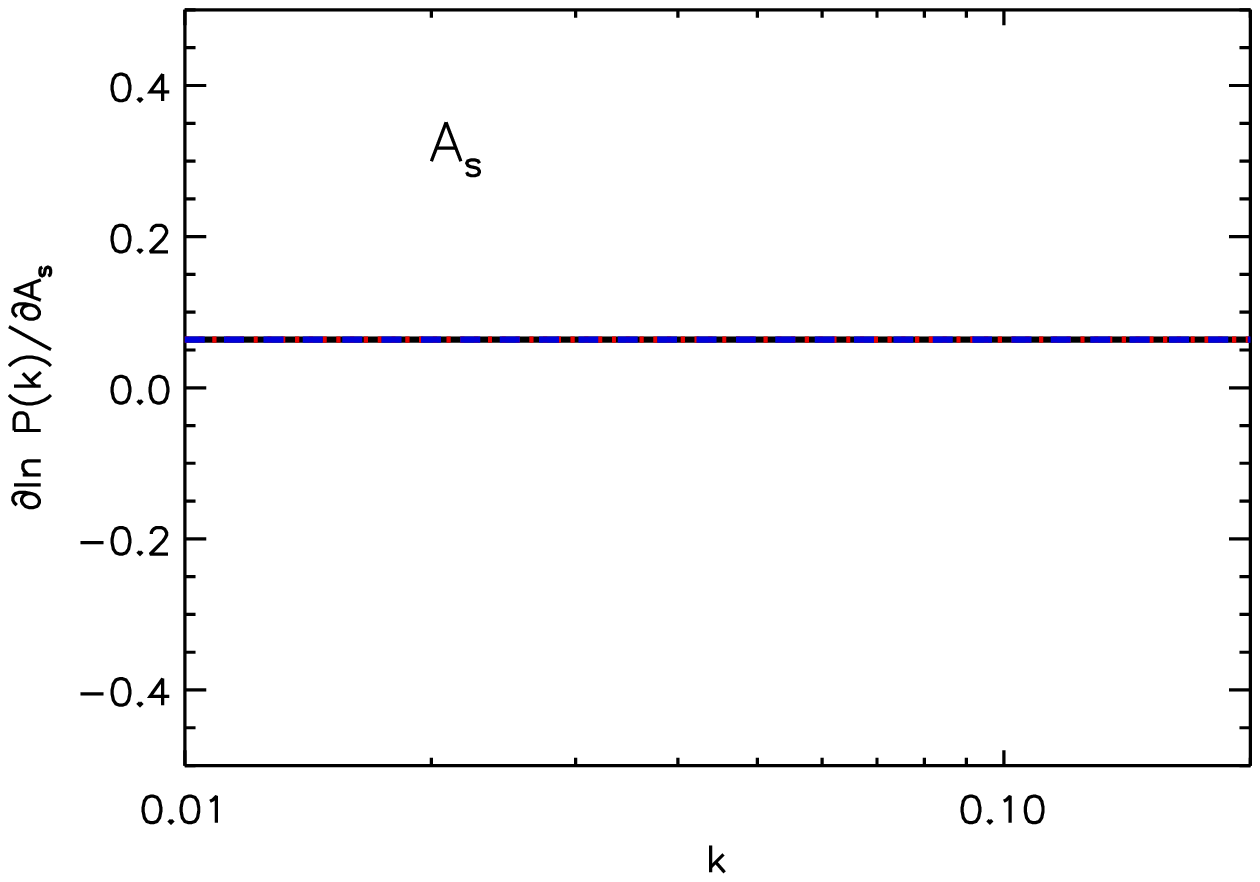}\\
 \includegraphics[trim = 0mm 10mm 0mm 10mm, scale=0.37, angle=0]{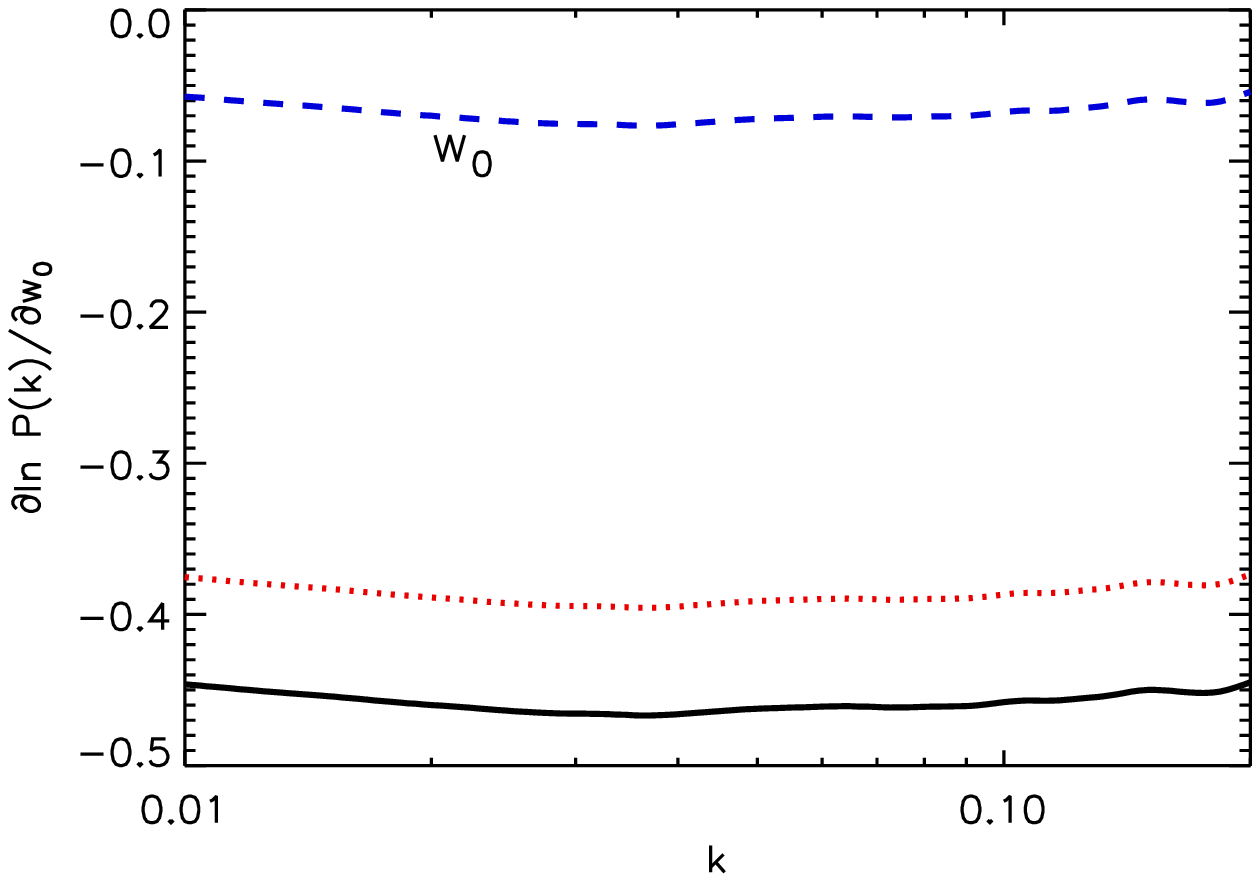} 
 & \includegraphics[trim = 0mm 10mm 0mm 10mm, scale=0.37, angle=0]{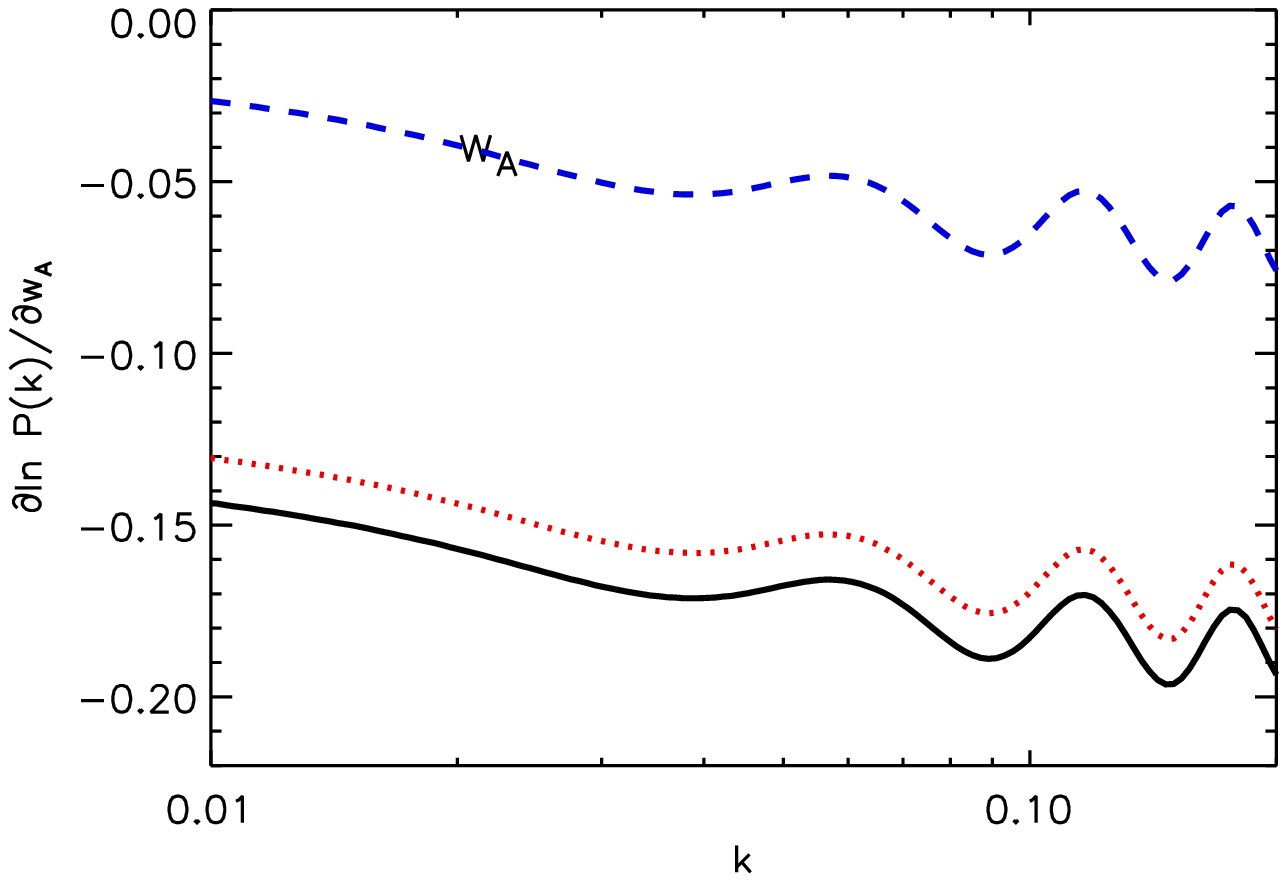}
 &
 \end{tabular}
 \end{center}
 \caption{Logarithmic derivatives of $P(k)$ with respect to the cosmological parameters for different redshifts: $z=0.35$ (solid black), $z=0.6$ (dotted red) and $z=3$ (dashed blue). An adiabatic fiducial model is assumed.}
 \label{fig:derivs}
 \end{figure*}

\begin{figure*}[ht!]
 \begin{center}
 \begin{tabular}{ccc}
 \includegraphics[trim = 0mm 0mm 0mm 20mm, scale=0.37, angle=0]{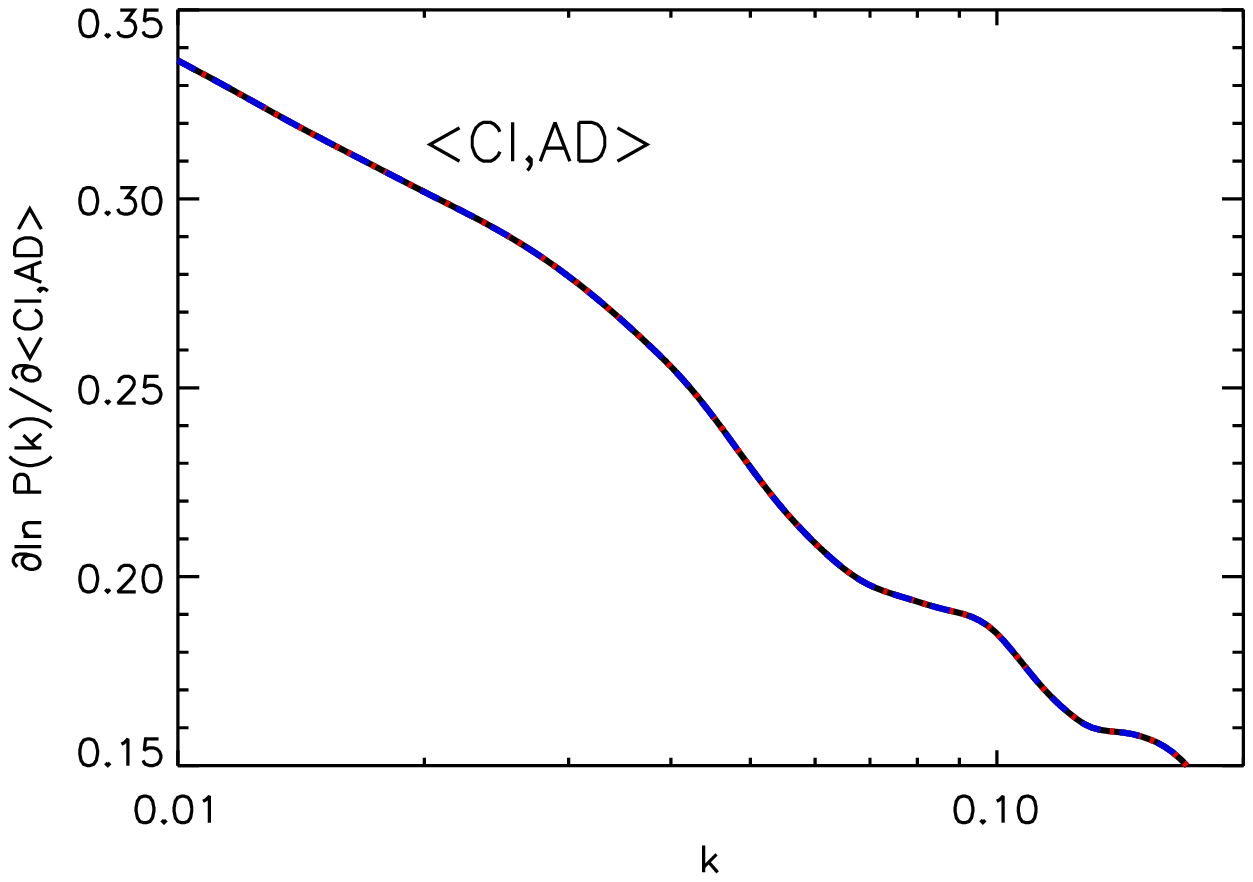} 
&\includegraphics[trim = 0mm 0mm 0mm 20mm, scale=0.37, angle=0]{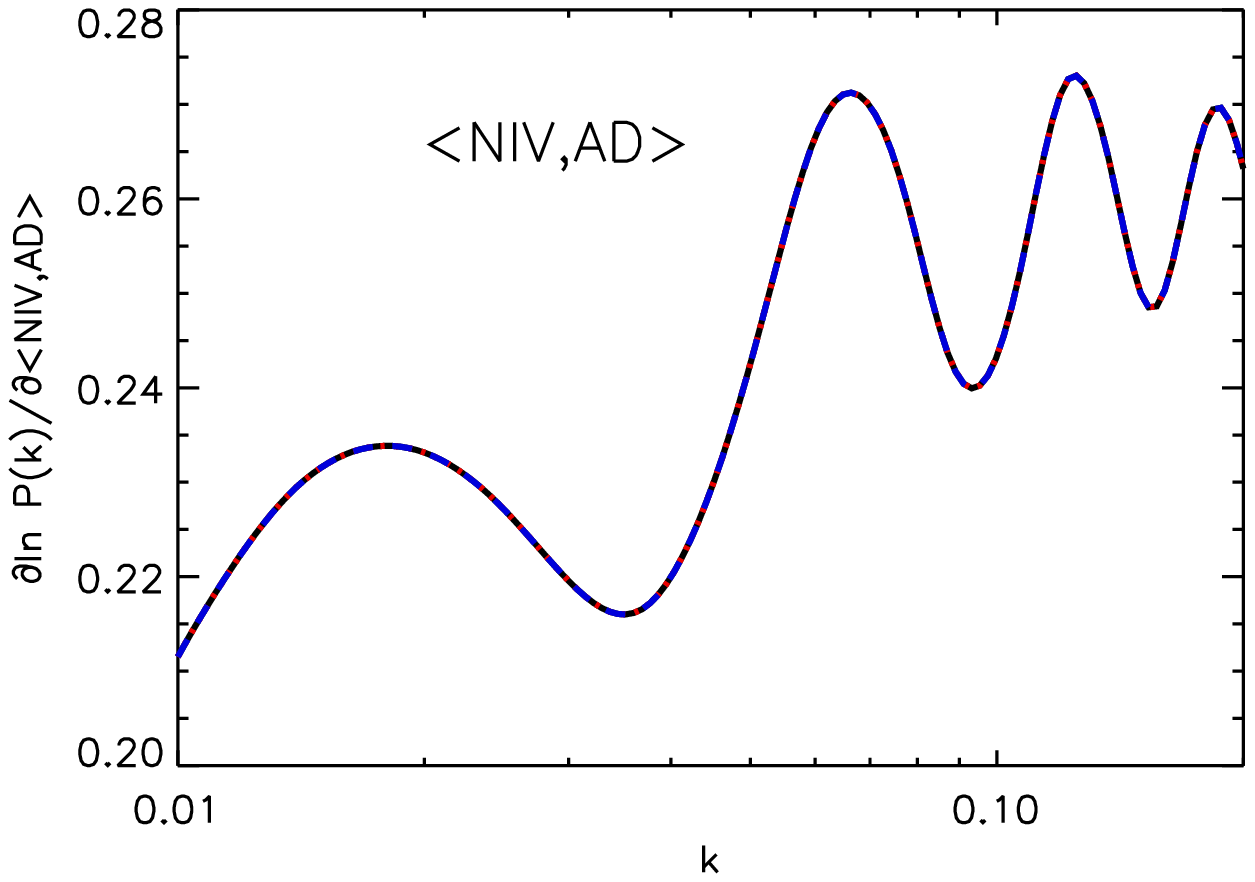}
&\includegraphics[trim = 0mm 0mm 0mm 10mm, scale=0.37, angle=0]{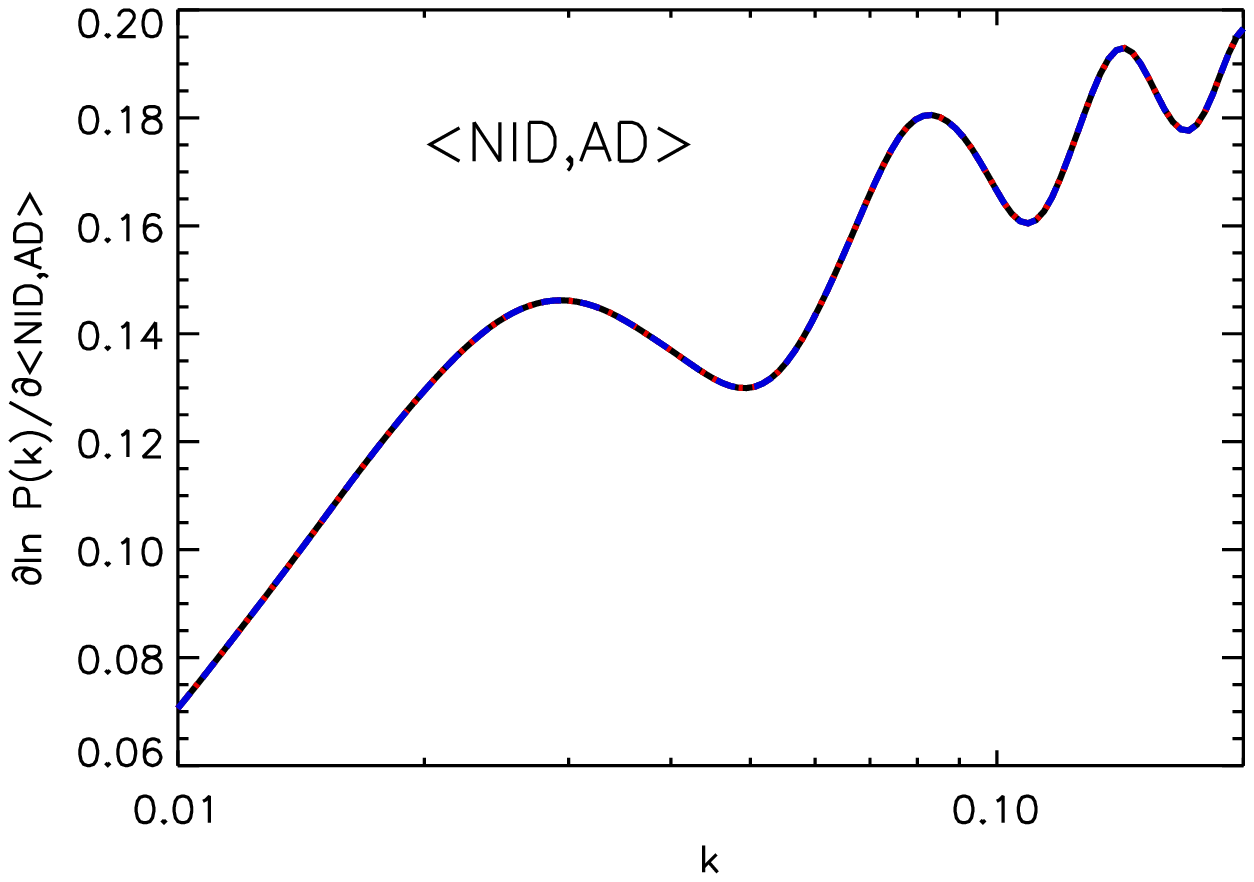}\\
 \includegraphics[trim = 0mm 0mm 0mm 10mm, scale=0.37, angle=0]{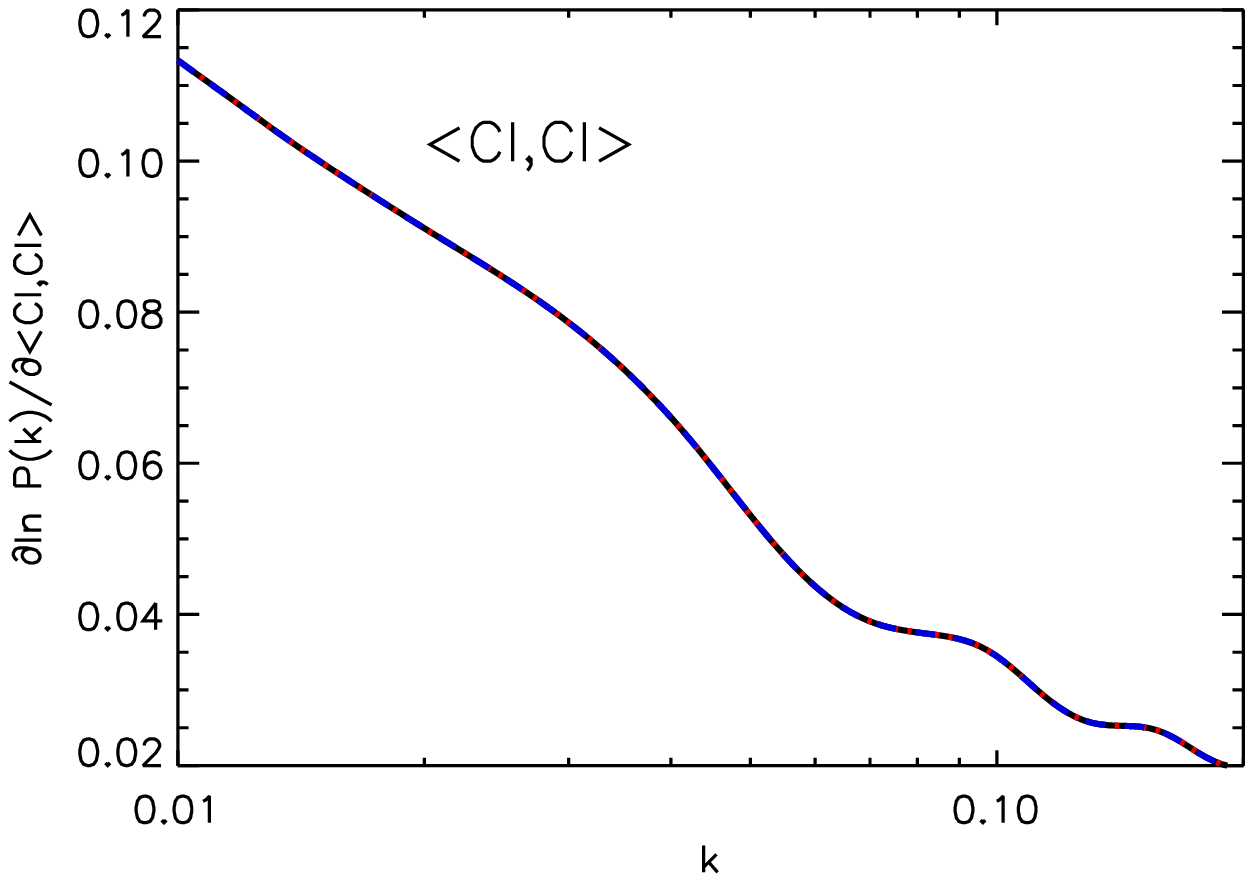} 
&\includegraphics[trim = 0mm 0mm 0mm 10mm, scale=0.37, angle=0]{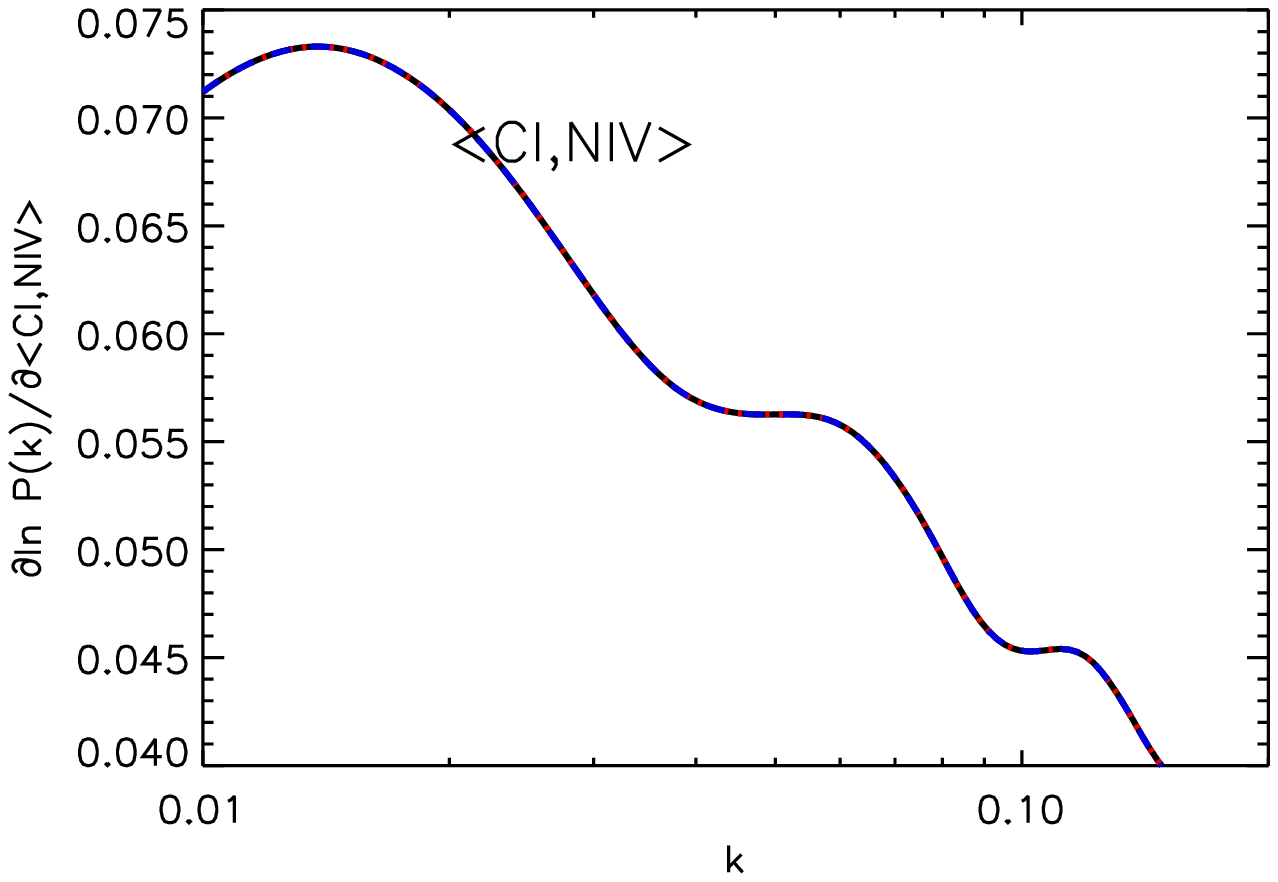}
&\includegraphics[trim = 0mm 0mm 0mm 10mm, scale=0.37, angle=0]{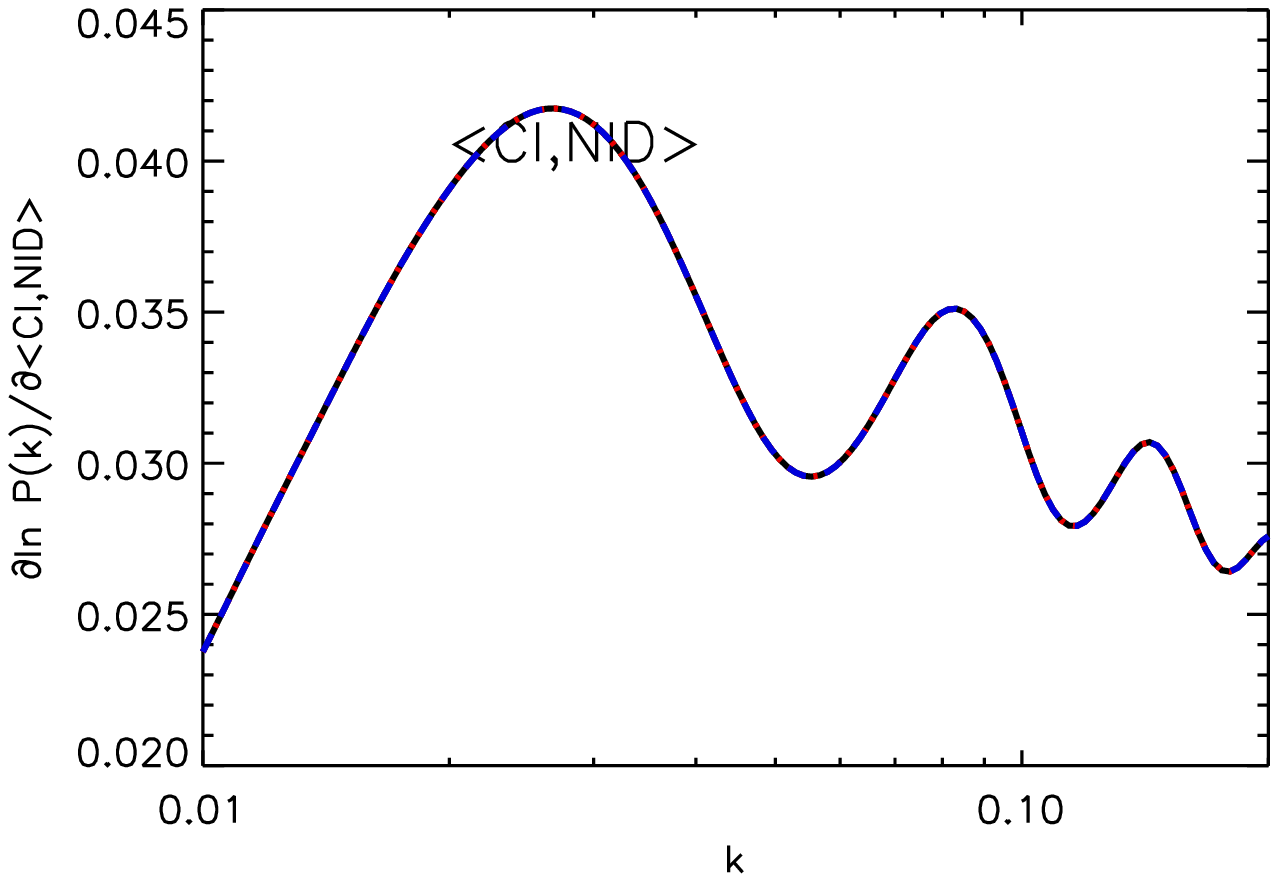}\\
 \includegraphics[trim = 0mm 10mm 0mm 10mm, scale=0.37, angle=0]{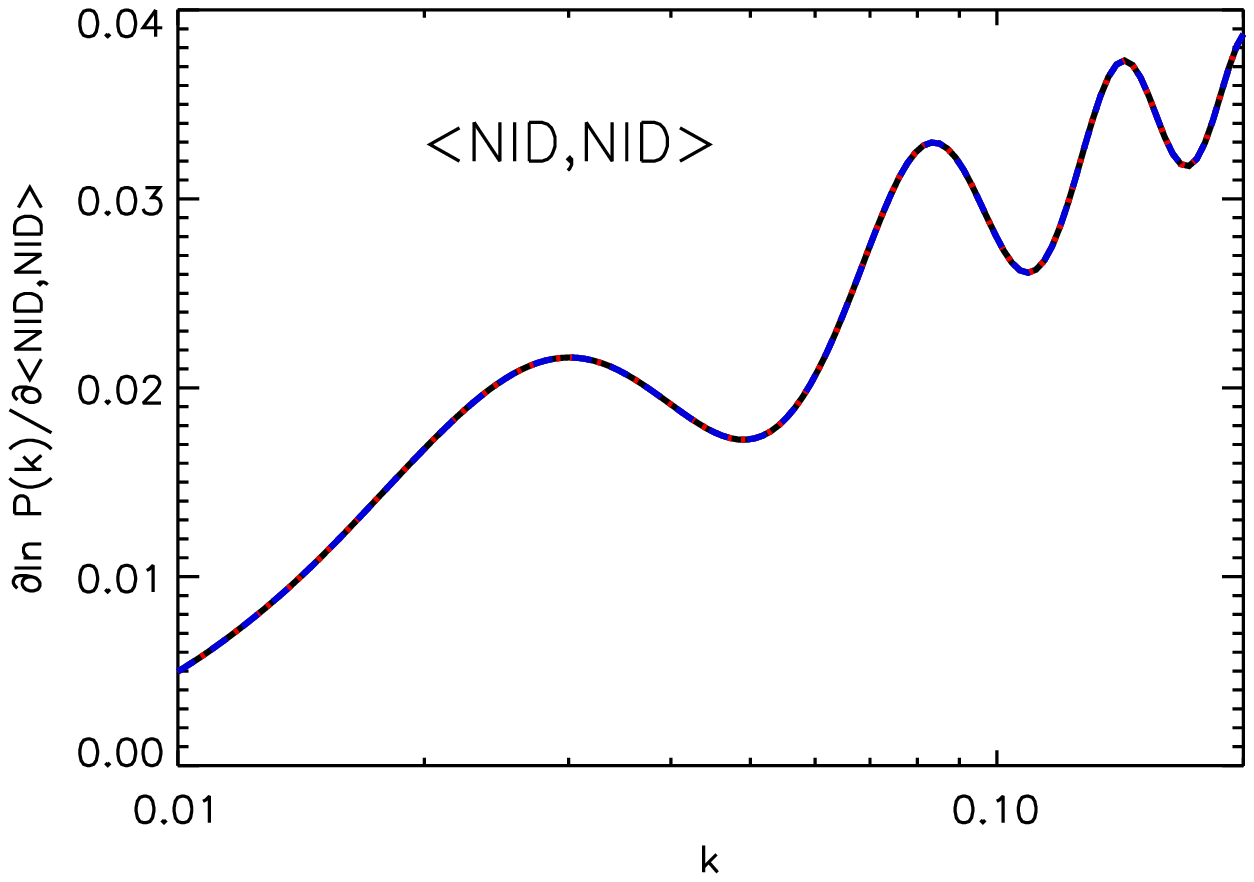}
&\includegraphics[trim = 0mm 10mm 0mm 10mm, scale=0.37, angle=0]{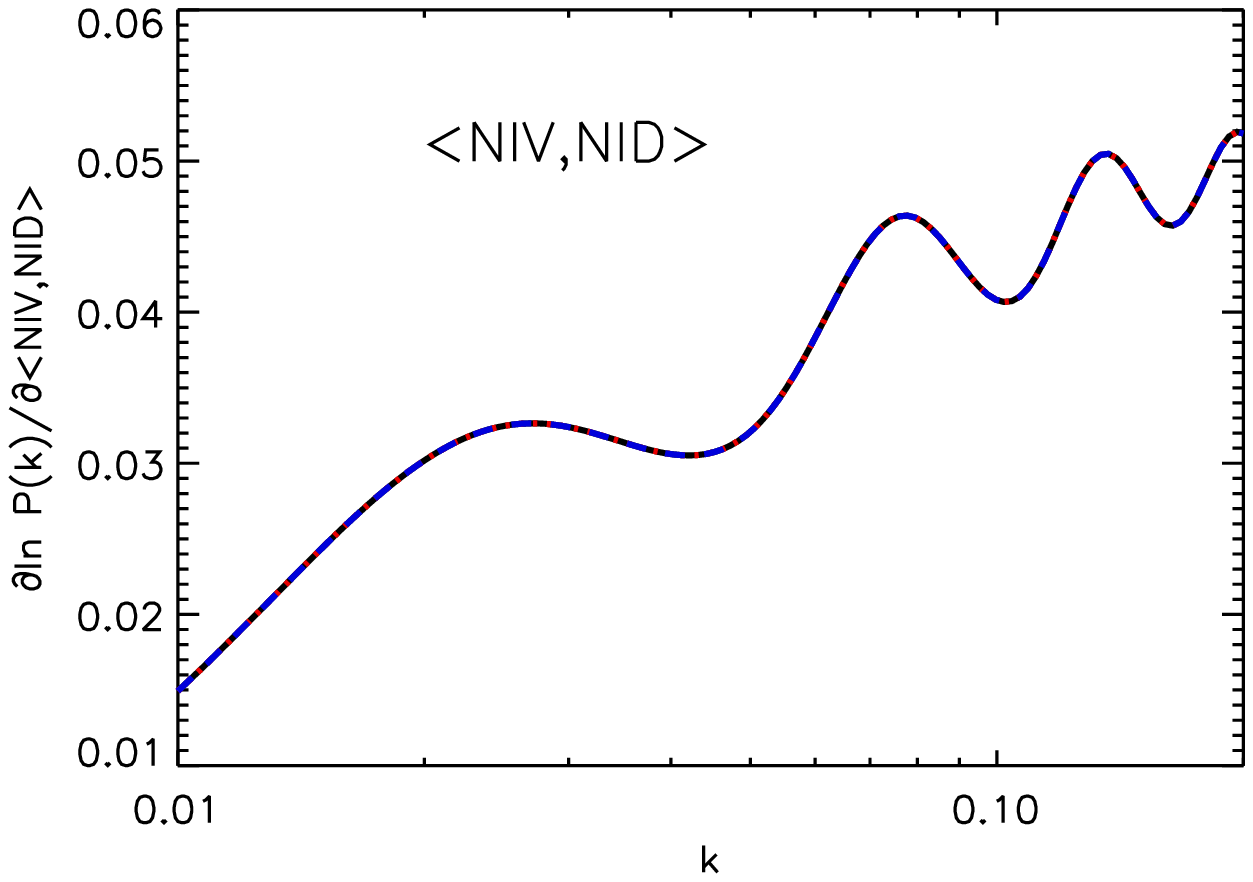}
&\includegraphics[trim = 0mm 10mm 0mm 0mm, scale=0.37, angle=0]{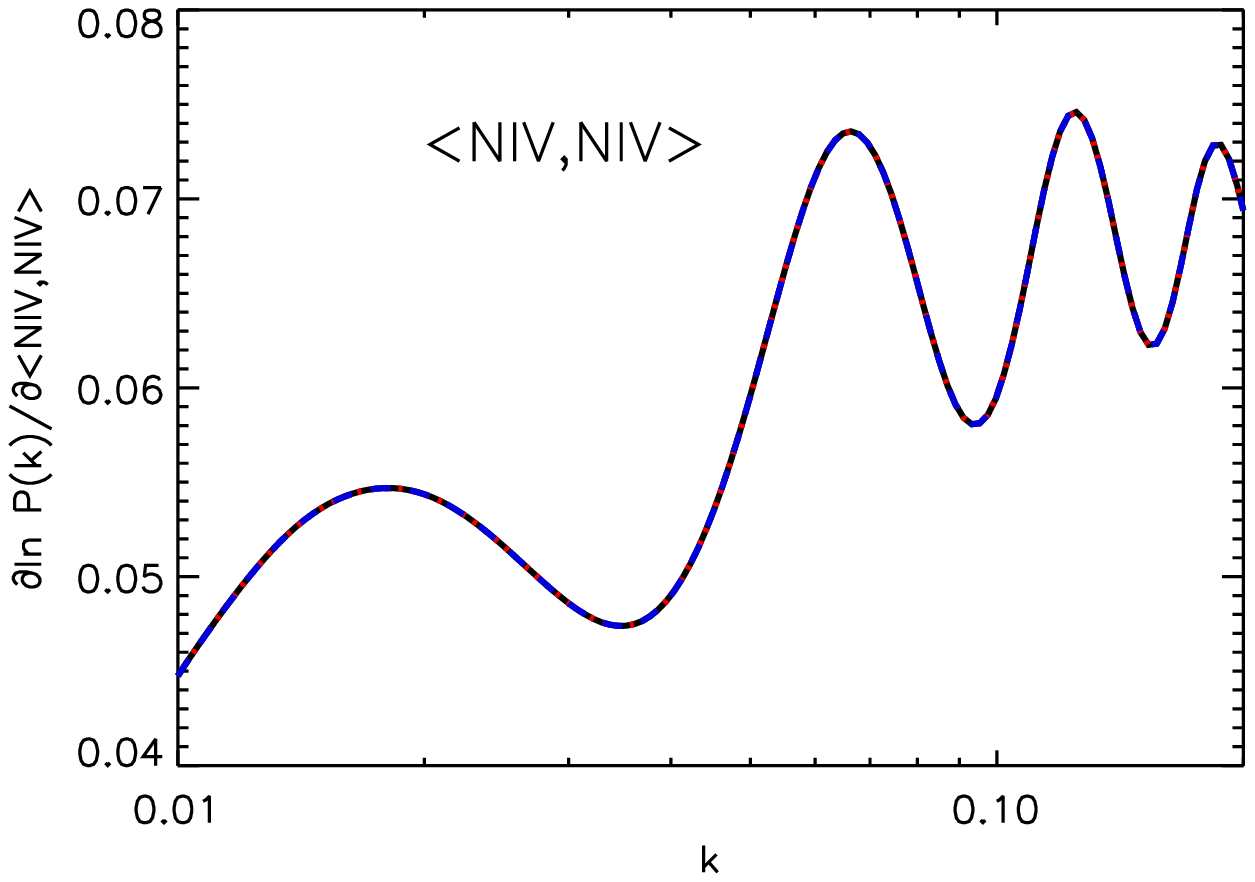}
 \end{tabular}
  \end{center}
\caption{Logarithmic derivatives of $P(k)$ with respect to the isocurvature parameters for different redshifts: $z=0.35$ (solid black), $z=0.6$ (dotted red) and $z=3$ (dashed blue). An adiabatic fiducial model is assumed.}
 \label{fig:derivs2}
 \end{figure*}
\clearpage

\end{document}